\documentclass[12pt]{article}
\usepackage{graphicx}

\usepackage[margin=1.1in]{geometry}

\usepackage{natbib}
\graphicspath{{Figures/}}

\title{New Tests of Randomness for Circular Data}
\author{Shriya Gehlot\\
    Operations and Decision Sciences\\ Indian Institute of Management Ahmedabad\\
        and \\
    Arnab Kumar Laha \\
    Operations and Decision Sciences\\ Indian Institute of Management Ahmedabad}
\date{}

\usepackage{amsmath}
\usepackage{amsthm,amssymb} 
\usepackage{comment} 
\usepackage{hyperref}
\usepackage{enumerate}

\usepackage{tikz}
\usetikzlibrary{arrows,petri,shapes,automata,positioning}

\usepackage{caption}        
\usepackage{subcaption}     
\usepackage{multirow} 

\usepackage{algorithm}      
\usepackage{algpseudocode}  
\usepackage{enumerate}      

\usepackage{apxproof} 

\theoremstyle{plain}
\newtheorem{theorem}{Theorem}[section]
\newtheorem{lemma}[theorem]{Lemma}
\newtheorem{proposition}[theorem]{Proposition}
\newtheorem{corollary}[theorem]{Corollary}

\theoremstyle{definition}   
\newtheorem{exmp}[theorem]{Example}

\newtheorem{remark}[theorem]{Remark}

\newcommand\numberthis{\addtocounter{equation}{1}\tag{\theequation}}

\def\spacingset#1{\renewcommand{\baselinestretch}%
{#1}\small\normalsize} \spacingset{1}
\spacingset{1.3}

\begin{document}

\maketitle

\begin{abstract}
    Randomness or mutual independence is an important underlying assumption for most widely used statistical methods for circular data. Verifying this assumption is essential to ensure the validity and reliability of the resulting inferences.  In this paper, we introduce two tests for assessing the randomness assumption in circular statistics, based on \textit{random circular arc graphs} (RCAGs). We define and analyze RCAGs in detail, showing that their key properties depend solely on the i.i.d. nature of the data and are independent of the particular underlying continuous circular distribution. Specifically, we derive the edge probability and vertex degree distribution of RCAGs under the randomness assumption. Using these results, we construct two tests: RCAG-EP, which is based on edge probability, and RCAG-DD, which relies on the vertex degree distribution. Through extensive simulations, we demonstrate that both tests are effective, with RCAG-DD often exhibiting higher power than RCAG-EP. Additionally, we explore several real-world applications where these tests can be useful.
\end{abstract}

\noindent%

\section{Introduction}
\label{sec:1}

One of the most common assumptions in data analysis is that a dataset is a `random sample' from the target population, which means that the observations are a realization of a set of mutually independent random variables.  The assumption is pervasive across many disciplines, including statistics, finance, economics, medicine, engineering, and other social sciences. For instance, in finance, stock returns and asset price changes are usually assumed to be random \citep{time_series}.  In economics, randomization is used to study the effects of interventions, such as monetary rewards for students on education in developing countries \citep{edu_Africa} or social interactions on group ties in an organization \citep{group_ties}. 
In medicine, the occurrence of genetic mutations is often modeled as random events \citep{bio_random}. In astronomy, meteoroid impacts are typically assumed to occur randomly in time and space \citep{fireball}. In statistics, many methodologies, including regression, goodness-of-fit tests, and one-sample t-test, rely on the assumption of random sampling.
Hence, it is crucial to check this assumption using a suitable statistical test before carrying out the data analysis. However, not many `randomness' tests are known in the literature, even for linear data.

In circular (directional) statistics, the data is expressed in the form of angles or directions. These angles could be represented as points on a unit circle by fixing an initial (zero) direction and an orientation.  The statistical treatment of circular data is considerably different from that of linear data, as it poses unique challenges. The reader may refer to foundational texts such as those by \cite{Batschelet}, \cite{Fisher}, and  \cite{SenGupta} for a better understanding of circular statistics.  
Recent advances in circular statistics have shown its application in various domains such as meteorology \citep{ocean,wind_dir},  ecology \citep{ecology1, ecology2}, biology \citep{bio1,bio2,bio3}, psychology \citep{psyc1}, social sciences \citep{so_sc1}, computer vision \citep{comp_vision1,comp_vision2}, and machine learning \citep{ml1,ml2}.

\begin{figure}[t]
	\centering
	\begin{subfigure}{.4\textwidth}
		\centering
		\includegraphics[scale=0.5]{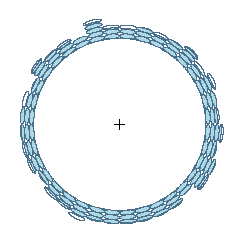}
		\caption{Uniform and Random}
	\end{subfigure}\hfil
	\begin{subfigure}{0.4\textwidth}
            \raggedleft
            \includegraphics[scale=0.47]{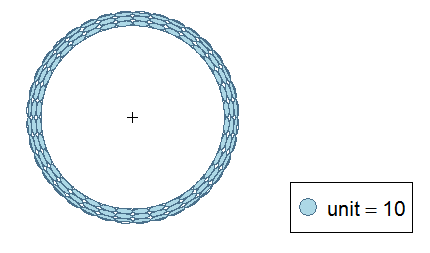}
		\caption{Uniform and Non-random}
	\end{subfigure}		
	\caption{Circular dot plot of data generated (a) randomly from the circular uniform distribution (b) non-randomly with points at a difference of $0.36^\circ$ each.}
	\label{fig1}
\end{figure}

The concepts of randomness and uniformity are often used interchangeably in circular statistics. However, they are not equivalent. If the observations are drawn `at random' from a circular uniform distribution, we expect to see no directional bias or one-sidedness when a circular dot plot is made using the observations (see Fig. \ref{fig1}a). Yet, it is possible to have circular uniformity without the observations being a realization of a set of mutually independent random variables (see Fig. \ref{fig1}b). Thus, a test of circular uniformity is not equivalent to a test of randomness for circular data.  While several tests have been proposed for circular uniformity in the existing literature, to the best of our knowledge, no test of randomness has been proposed. Additionally, it is important to remember that a random sample of circular observations can follow a distribution other than the circular uniform distribution and tests for checking randomness are required for these situations too.   

To exemplify the importance of testing randomness, consider the study \cite{fireball}, in which the authors try to find the directional distribution of location (latitude and longitude) of extraterrestrial objects like meteor falls, fireball showers, and craters. Usually, it is assumed that meteoroid impacts are random in time and space, meaning the impacts are uniformly spread throughout the year and on the surface of the Earth \citep{laFuenteMarcos}. 
However, the authors question the uniformity of location and explore the suitability of the von Mises distribution to describe these events using  Rao’s spacing tests \citep{rao's_spacing} and Watson’s test \citep{Watson}, respectively. Both these tests rely on the assumption that the data is a random sample from an underlying distribution (i.e., i.i.d.). The results suggest that the meteor falls and fireball showers are non-uniform but do not follow the von Mises distribution. However, was the failure to fit the von Mises distribution due to the distribution itself or because the data was not truly random? The authors did not test for independence among observations, leaving open the possibility that non-randomness affected their findings. We will dive deeper into this case in Section \ref{real_data} of this paper.

Testing for randomness aims to uncover possible dependence among the observations. Dependence can take many forms, making an omnibus test for dependence challenging. 
Most randomness tests for linear data focus on identifying specific types of dependence.  
This challenge becomes even more pronounced when working with directional data. As a result, developing a randomness test for circular data that can accommodate diverse scenarios is not easy. 

Over the decades, researchers have developed various robust tests to check for uniformity in the circular data. The most commonly used uniformity tests include \cite{Kuiper}, \cite{Ajne}, \cite{Rayleigh}, Watson, and Rao’s spacing tests. \cite{GPG} gives a comprehensive overview of these tests. Beyond uniformity, researchers have also explored independence in bivariate circular data using correlation measures \citep{cor1, cor2, cor3, cor4} and other methods \citep{bivar1, bivar2, bivar3, bivar4, bivar5}. Despite these advances, to the best of our knowledge, there is no test available for checking randomness in univariate circular data. Such a test is crucial for detecting dependencies inherent in circular data, ensuring accurate analysis and inference with circular datasets. 

In this paper, we propose a new approach to developing tests for checking randomness in circular data.  For this, we introduce the concept of \textit{random circular arc graphs (RCAGs)} for circular data, analogous to \textit{random interval graphs (RIGs)} for linear data. We systematically investigate several key properties of RCAGs, including edge probability, vertex degree distribution, maximum and minimum degrees, connectedness, and the existence of Hamiltonian cycles. Importantly, we establish that these properties are invariant to the choice of distribution and rely solely on the independence of the observations. Building on this critical finding, we construct new randomness tests suitable for a wide range of circular data applications. Our results demonstrate that the proposed tests provide a versatile and effective tool for identifying randomness across a wide range of scenarios.

This paper is organised as follows. Section \ref{RCAG} provides an overview of RIGs and \textit{circular arc graphs (CAGs)}, laying the foundation for the introduction of RCAGs. In this section, we also explore various properties of RCAGs that form the basis for developing our proposed randomness tests. In Section \ref{ToR}, we present two new tests to check for randomness in circular data. In Section \ref{EP_algo}, we use the edge probability of the RCAG to introduce the  \textit{RCAG-Edge Probability} test. In Section \ref{DD_algo}, we use the vertex degree distribution of RCAG to introduce the  \textit{RCAG-Degree Distribution} test to assess randomness. Section \ref{exp} presents the findings from an extensive simulation-based performance evaluation of these tests across various setups. In Section \ref{real_data}, we apply our tests to two real-life circular datasets: a wind-direction dataset and a dataset on the locations of extraterrestrial objects, such as meteor falls. Section \ref{conclusions} provides a summary of the findings and highlights our key observations.

\section{Random Circular Arc Graphs}
\label{RCAG}

\subsection{Introduction and Basic Definitions}
\label{intro}


In this section, we begin by providing the definitions of some of the terms that will be extensively used in this article. Let us consider a finite collection of sets $s_1, s_2, \ldots, s_n$ and let  $S = \{s_1, s_2, \ldots, s_n\}$. The graph $G = (V, E)$ is referred to as the \textit{intersection graph} of the sets in $S$ if  $V = V(G) = \{s_1, s_2, \ldots, s_n\}$, and $\{s_i, s_j\} \in E = E(G)$ if $s_i \cap s_j \ne \emptyset, i \neq j$. In other words, an edge exists between any two vertices if their corresponding sets have a non-empty intersection. When the sets are intervals on the real line (or any linearly ordered set), the intersection graphs derived from these intervals are specifically termed \textit{interval graphs} \citep{2}.

In a random graph, the presence of edges between vertices is determined by a random process. We now define a process for the construction of random interval graphs as given in \cite{3}, which has proved to be useful for studying the properties of these graphs.  Later,  we extend these ideas to the circular case.

Let $X_1,X_2,\dots, X_n, Y_1,Y_2,\dots,Y_n$ be $2n$ independent random variables that all follow the same continuous distribution $F$ (cdf), i.e., these random variables are i.i.d. $F$. Then $G=(V,E)$ is the \textit{random interval graph} (RIG) corresponding to this sequence if the $i^{th}$ vertex in $G$ corresponds to the interval $[X_i,Y_i]$ if $X_i < Y_i$ or $[Y_i,X_i]$ if $Y_i < X_i$. Here, two vertices are adjacent (i.e., there is an edge joining these vertices) when the corresponding intervals intersect. Since the properties of RIGs are distribution invariant, without loss of generality, it can be assumed that $X_1, X_2,\dots, X_n, Y_1, Y_2,\dots, Y_n$ are i.i.d. Uniform[0,1] \citep{3, 4}.

Extending the concept of an interval graph, a \textit{circular arc graph} (CAG) is defined by taking the sets $s_i, i=1, 2, \ldots, n$ to be the arcs on a unit circle. The intersection graph formed by these arcs is referred to as a circular arc graph \citep{2}.

Circular arc graphs have been extensively studied in the existing literature (see \cite{CAG_S1} and \cite{CAG_S2} for recent surveys on it). However, to the best of our knowledge, the scenario in which arcs (or vertices) are randomly generated in such graphs has not been explored.  We now extend the definition of RIG to the circular case and examine the distinctions that arise when compared to the linear scenario. 
Let $\Theta_1,\Theta_2,\dots, \Theta_n, \Phi_1,\Phi_2,\dots,\Phi_n$ be $2n$ independent random variables with continuous circular distribution $F$.  Then $G=(V,E)$ is the \textit{random circular arc graph} (RCAG) corresponding to this sequence if the $i^{th}$ vertex in $G$ corresponds to the arc starting at angle $\Theta_i$ and ending at angle $\Phi_i$ traversed in an anticlockwise manner. Here, two vertices are adjacent when the corresponding arcs intersect. 

\begin{figure}[t]
    \centering
    \subfloat[]{
        \begin{tikzpicture}[scale=0.5]
            \draw[color=red, thick] (0,0) --(6,0);
            \draw[color= blue, thick] (0,-0.3) -- (2,-0.3);
            \draw [color = cyan, thick] (4,-0.3) -- (6,-0.3);
            \draw [red, thick] (0,0) circle(.8pt) node[above] {1};
            \draw [red, thick] (2,0) circle(.8pt) node[above] {2};
            \draw [red, thick] (4,0) circle(.8pt) node[above] {3};
            \draw [red, thick] (6,0) circle(.8pt) node[above] {4};
            \draw [blue, thick] (0,-0.3) circle(.8pt);
            \draw [blue, thick] (2,-0.3) circle(.8pt);
            \draw [cyan, thick] (4,-0.3) circle(.8pt);
            \draw [cyan, thick] (6,-0.3) circle(.8pt);
            \node[blue] at (1,-0.8) {$I_1$};
            \node[cyan] at (5,-0.8) {$I_2$};
        \end{tikzpicture}
    }
    \hfill
    \subfloat[]{
        \begin{tikzpicture}[scale=0.5]
            \draw[color=red, thick] (0,0) --(6,0);
            \draw[color= blue, thick] (0,-0.3) -- (2,-0.3);
            \draw [color = cyan, thick] (4,-0.3) -- (6,-0.3);
            \draw [red, thick] (0,0) circle(.8pt) node[above] {1};
            \draw [red, thick] (2,0) circle(.8pt) node[above] {2};
            \draw [red, thick] (4,0) circle(.8pt) node[above] {3};
            \draw [red, thick] (6,0) circle(.8pt) node[above] {4};
            \draw [blue, thick] (0,-0.3) circle(.8pt);
            \draw [blue, thick] (2,-0.3) circle(.8pt);
            \draw [cyan, thick] (4,-0.3) circle(.8pt);
            \draw [cyan, thick] (6,-0.3) circle(.8pt);
            \node[blue] at (1,-0.8) {$I_1$};
            \node[cyan] at (5,-0.8) {$I_2$};
        \end{tikzpicture}
    }
    \hfill
    \subfloat[]{
        \begin{tikzpicture}[scale=0.4]
            \def \n {4}
            \def \radius {1.8}
            \draw [red, thick] (0,0) circle [radius=1.8]
            foreach\s in{1,...,\n}{
                (360/\n*\s + 90:-\radius) circle(.4pt)circle(.8pt)circle(1.2pt)
                node[anchor=360/\n*\s+90]{\tiny{$\s$}}
            };
            \draw[thick, color=blue]  (0,2.236) arc[start angle=90, end angle=0,radius=2.236cm] node[midway,sloped,above] {$A_1$};
            \draw[thick,color=cyan] (-2.236,0) arc[start angle=180, end angle= 270,radius=2.236cm] node[midway,sloped,below] {$A_2$} ;
        \end{tikzpicture}
    }
    \hfill
    \subfloat[]{
        \begin{tikzpicture}[scale=0.4]
            \def \n {4}
            \def \radius {1.8}
            \draw [red, thick] (0,0) circle [radius=1.8]
            foreach\s in{1,...,\n}{
                (360/\n*\s + 90:-\radius) circle(.4pt)circle(.8pt)circle(1.2pt)
                node[anchor=360/\n*\s+90]{\tiny{$\s$}}
            };
            \draw[thick, color=blue]  (0,2) arc[start angle=90, end angle=0,radius=2cm] node[midway,sloped,below] {$A_1$};
            \draw[thick,color=cyan] (-2.236,0) arc[start angle=180, end angle=-90,radius=2.236cm] node[midway,sloped,above] {$A_2$} ;
        \end{tikzpicture}
    }
    \caption{Intervals and arcs illustrating the effect of order change in the circular case.}
    \label{fig2}
\end{figure}
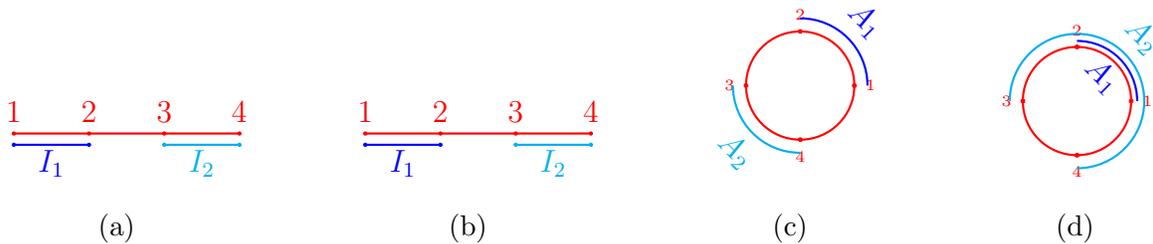

We elucidate the distinction between linear and circular scenarios through an illustrative example involving four random numbers, as portrayed in Fig. \ref{fig2}. Initially, adopting the sequence $X_1= \Theta_1 =1, X_2 = \Theta_2 =3, Y_1= \Phi_1 = 2, Y_2= \Phi_2 = 4$, we observe the same intervals and arcs: $I_1 = A_1 = [1, 2]$ and $I_2 = A_2 = [3, 4]$.  Now, altering the order, say to $X_1= \Theta_1 =1, X_2= \Theta_2 = 4, Y_1= \Phi_1 =2, Y_2= \Phi_1 = 3$, produces the same intervals as before ($I_1 = [1, 2]$ and $I_2 = [3, 4]$), yet the different arcs ($A_1 = [1, 2]$ and $A_2 = [4, 3]$). Consequently, these sequences generate identical RIGs but different RCAGs. This property, that the order of numbers affects the arcs formed, is the primary reason for the differences observed in the properties of RCAG compared to RIG.  

The significance of RIG and RCAG lies in the fact that the properties of these random graphs depend only on whether the observations are randomly generated, not on their distribution. Consequently, without loss of generality, we can assume $F$ to be Circular Uniform when analyzing the properties of RCAGs. The properties of RIGs have already been studied in literature by \cite{3}. 
In this paper, we focus on analyzing the properties of RCAGs, which will serve as the foundation for developing our tests of randomness for circular data. 


\subsection{Edge probability}
\label{edge_prob_sec}


The first property we examine is the probability of an edge existing between two vertices in an RCAG.

\begin{lemma} 
    Let $\Theta_i,\Phi_i, 1\leq i \leq n$ be $2n$ independent random variables which follow the circular uniform distribution. Then,
    in the RCAG formed by $\Theta_i's$ and $\Phi_i's$, the probability of an edge existing between any two randomly chosen vertices is \(\frac{5}{6}\).
    \label{edge_prob}
\end{lemma}

\begin{proof}
    The probability of an edge between any two vertices of an RCAG is equal to the probability of the intersection of two arcs on a unit circle.
    Let $[\theta_1,\phi_1]$ and $[\theta_2,\phi_2]$ be two arcs. Then they do not intersect in either of the following conditions:
(i) $\theta_1 < \phi_1 < \theta_2 < \phi_2,$ (ii) $\phi_2 < \theta_1 < \phi_1 < \theta_2$, (iii) $ \theta_2 < \phi_2 < \theta_1 < \phi_1$, (iv) $\phi_1 < \theta_2 < \phi_2 < \theta_1$.
Therefore, the probability that two arcs are non-intersecting is $4/4! = 1/6$. Hence the result.
\end{proof}

By contrast, \cite{3} demonstrates that this probability is $2/3$ for an RIG. This result highlights the increase in intersection probability from $2/3$ for intervals in an RIG to $5/6$ for arcs in an RCAG.  The higher probability in the circular case arises because the order of numbers within an arc matters, unlike in linear intervals as depicted in Fig. \ref{fig2}.

Next, we extend this result to any circular distribution with the pdf $f(\cdot)$ and the cdf $F(\cdot)$. Here $\theta_i,\phi_i$’s are i.i.d. by the construction of the RCAG.

\begin{lemma}
    The probability of an edge between two vertices in an RCAG is independent of the choice of circular distribution $F$. 
\label{edge_prob_general}
\end{lemma}

\begin{proof}
    Let $[\theta_1,\phi_1]$ and $[\theta_2,\phi_2]$ be two arcs of an RCAG. Then, they are non-intersecting for four cases as given in Lemma \ref{edge_prob}. Now, if we take any one of the four cases, say $\theta_1 < \phi_1 < \theta_2 < \phi_2$, then
\begin{align*}
    P(\theta_1 < \phi_1 < \theta_3 < \phi_2) & = \int_{0}^{2\pi} \int_{\theta_1}^{2\pi} \int_{\phi_1}^{2\pi} \int_{\theta_2}^{2\pi} f(\theta_1) f(\phi_1) f(\theta_2) f(\phi_2) d\phi_2 d\theta_2 d\phi_1 d\theta_1\\
    & = \int_{0}^{2\pi} \int_{\theta_1}^{2\pi} \int_{\phi_1}^{2\pi} f(\theta_1) f(\phi_1) f(\theta_2) [F(2\pi) - F(\theta_2)] d\theta_2 d\phi_1 d\theta_1\\
    & =  \frac{1}{24}
\end{align*}
by applying standard integration techniques. Similarly, the probability for the other three cases will also be $1/24$. These probabilities are the same as ones obtained for circular uniform distribution as shown in Lemma \ref{edge_prob},  giving us the required result.
\end{proof}

Lemmas \ref{edge_prob} and \ref{edge_prob_general} together will help us propose a test for randomness as described in Section \ref{EP_algo}.


\subsection{Number of Edges}
\label{no._edge}


In this section, we examine the fundamental characteristics of the edges in the RCAG, which lay the groundwork for exploring more advanced structural properties in subsequent sections. 
Let $X_{ij}$ =1 if there is an edge between vertices $i$ and $j$ ($i \ne j$) and $X_{ij}$ =0  otherwise. Then, by Section \ref{edge_prob_sec} $X_{ij} \sim Bernoulli (5/6)$ for RCAGs. 

\begin{lemma}
    \label{lemma_edge_dept}
    Let $X_{ij}$ be as defined above. Suppose $\{i,j\} \cap \{k,l\} \ne \emptyset$. Then $X_{ij}$ is not independent of $X_{kl}$.
\end{lemma}

\begin{proof}
    Consider three arcs $A_1=[\theta_1,\phi_1], A_2=[\theta_2,\phi_2],$ and $A_3=[\theta_3,\phi_3]$.
Then as described in Lemma \ref{edge_prob}, $X_{12}=0 $ when (i) $\theta_1 < \phi_1 < \theta_2 < \phi_2,$ (ii) $\phi_2 < \theta_1 < \phi_1 < \theta_2$, (iii) $ \theta_2 < \phi_2 < \theta_1 < \phi_1$, (iv) $\phi_1 < \theta_2 < \phi_2 < \theta_1$ and 
$X_{13}=0$ when (a) $\theta_1 < \phi_1 < \theta_3 < \phi_3,$ (b) $\phi_3 < \theta_1 < \phi_1 < \theta_3$, (c) $ \theta_3 < \phi_3 < \theta_1 < \phi_1$, (d) $\phi_1 < \theta_3< \phi_3 < \theta_1$. Then, the number of ways in which both $X_{12}=0 $ and $X_{13}=0$ is equal to arranging six numbers $\theta_1,\phi_1,\theta_2,\phi_2,\theta_3,\phi_3$ such that one of the four conditions for both $X_{12}=0 $ and $X_{13}=0$ is satisfied. The number of ways in which (i) and (a) hold together is 6, (i) and (b) is 3, (i) and (c) is 1, (ii) and (a) is 3, (ii) and (b) is 4, (ii) and (c) is 3, (iii) and (a) is 1, (iii) and (b) is 3, (iii) and (c) is 6 and (iv) and (d) is 6. Therefore, $P(X_{12}=0\cap X_{13}=0) = \frac{36}{6!} = \frac{1}{20}$ which is greater than $P(X_{12}=0)P(X_{13}=0) = \frac{1}{6}\cdot\frac{1}{6} = \frac{1}{36}$.
\end{proof}

Lemma \ref{lemma_edge_dept} implies that while the edge variables $X_{ij}, i\neq j$ are identically distributed,  they are not mutually independent. This dependence introduces challenges in analyzing more complex properties of RCAGs and in constructing tests for randomness, as we will explore in later sections.

We now turn to analyzing the total number of edges in an RCAG.
Let $\mathcal{G}_n$ denote the set of all possible RCAGs with $n$ vertices. 

\begin{theorem}
\label{no_edges_thm}
   Let $G \in \mathcal{G}_n$ and $X$ be the total number of edges in $G$. Then $ \qquad$ $ P\left(|X-E(X)| \geq cn^{\frac{7}{4}} \right) \rightarrow 0$ as $n \rightarrow \infty$ for some small $c > 0$, where $E(X) = \frac{5}{6}n^2 + o(n^2)$. In other words,
   $G$ have $\frac{5}{6}n^2 + o(n^2)$ edges. 
\end{theorem}

\begin{proof}
    For $G \in \mathcal{G}_n$ and $1 \leq i,j \leq n$, let $X_{ij} = 1$ if $i \sim j$ and $X_{ij} = 0$ otherwise. Then $P(X_{ij} = 1) = 5/6$ by Lemma \ref{edge_prob}. Let $X = \sum_{(i,j) \in S} X_{ij}$ be the total number of edges in $G$. Then
$$E(X) =\sum_{ (i,j) \in S} E(X_{ij}) = \frac{5}{6} \times 2 \binom{n}{2} = \frac{5}{6} n^2 -\frac{5}{6} n  = \frac{5}{6} n^2 + o(n^2)$$
since  $E(X_{ij})=\frac{5}{6}$ and there are $2 \binom{n}{2}$  terms in the summation.  
Now, $Var(X) = E(X^2) - E(X)^2$ where
$$E(X^2) = E\left[ \sum_{(i,j) \in S} X_{ij} \sum_{(k,l) \in S} X_{kl} \right].$$
The right-hand side can be partitioned into two scenarios. First, when all $i$, $j$, $k$, and $l$ are distinct, and second, when there is an overlap between the sets $\{i,j\}$ and $\{k,l\}$.
Then 
$$E(X^2) = \sum_{\substack {(i,j) \in S \\ (k,l) \in S \\ \{i,j\} \cap \{ k, l\} = \emptyset}} E [ X_{ij}X_{kl}] + \sum_{ \substack{i,j) \in S \\  (k,l) \in S \\ \{i,j\} \cap \{ k, l\} \ne \emptyset}} E \left[ X_{ij}X_{kl} \right ].$$
For the first case, the number of ways of choosing $i,j,k,l$ will be  $n(n-1)(n-2)(n-3)$, and since all four vertices are distinct, the event $X_{ij}$ is independent of the event $X_{kl}$.  Therefore, each of these $E(X_{ij}X_{kl}) = E(X_{ij})E(X_{kl}) = 25/36$. For the second case, since there is an overlap between the sets $\{i,j\}$ and $\{k,l\}$, this means we are choosing either (i) two or (ii) three distinct numbers, and the remaining numbers take values same as that of $i$ or $j$. 
For (i), $k$ and $l$ take same values as $i$ and $j$, which could happen in $n(n-1)\cdot2 \cdot 1$ possible ways. For (ii), one of $k$ or $l$ takes same values as $i$ or $j$, which could happen in $2 \cdot n(n-1)(n-2)\cdot 2$ possible ways.
Therefore, we will have $2n(n-1)(2n -3)$ terms in the summation.
Also, since $X_{ij}, X_{kl} \in \{0,1\}$, the contribution of $E(X_{ij}X_{kl})$ from each of these terms will be less than or equal to 1. Therefore,
\begin{align*}
    E(X^2) &\leq \frac{25}{36} n(n-1)(n-2)(n-3) + 2n(n-1)(2n -3)\\
    & = n(n-1) \left(\frac{25}{36}n^2 + \frac{19}{36}n - \frac{11}{6} \right)
\end{align*}
giving
\begin{align*}
    Var(X) 
    & \leq n(n-1) \left(  \frac{44n}{36} - \frac{11}{6}\right ) = O(n^3)
\end{align*}
Then, by Chebyshev's inequality,
$$P\left(|X-E(X)| \geq cn^{\frac{7}{4}} \right) \leq \frac{Var(X)}{c^2n^{7/2}} \rightarrow 0, $$
as $n\rightarrow \infty$ for any small $c>0$.
\end{proof}


\subsection{Degree of a vertex}
\label{degree}


We now shift our focus to understanding how the degree of a vertex behaves in an RCAG. Due to the inherent dependence among edges, the degrees of the vertices are also interdependent, making it challenging to derive a general distribution function for vertex degrees. To address this, we first compute the probability that a fixed vertex is connected to any other vertex in the RCAG and then use it to derive an expression for the vertex degree distribution.

\begin{proposition}
\label{fixed_arc_prob}
    Let $A_1 = [\theta_1,\phi_1]$ be a fixed arc of $G \in \mathcal{G}_n$. Let $X_i = 1$ if $A_i \cap A_1 \neq \emptyset$ for $ 2 \leq i \leq n$, and $X_i = 0$ otherwise. Then
    $$ P(X_i = 0) = \frac{1}{2(2\pi)^2} \left((\theta_1 - \phi_1) \mod 2\pi \right)^2. $$
\end{proposition}

\begin{proof}
    Note that the event  $X_i=0$ occurs if $A_i \equiv [\theta_i,\phi_i]$ does not intersect with $A_1$. This happens when either of the following conditions is satisfied: 
\begin{enumerate}
    \item $\theta_i < \phi_i < \theta_1$ or $\phi_1 < \theta_i < \phi_i$ or $\phi_i < \theta_1 < \phi_1 < \theta_i$ for $\theta_1 \leq \phi_1 $,
    \item $\phi_1 < \theta_i < \phi_i < \theta_1$ for $\phi_1 \leq \theta_1$.
\end{enumerate}
Now, $P(\theta_i < \phi_i < \theta_1 < \phi_1) = \int_{0}^{\theta_1} \int_{\theta_i}^{\theta_1} \left( \frac{1}{2\pi} \right)^2 d\phi_i d\theta_i  
= \frac{1} {2} \left (\frac{\theta_1}{2\pi} \right)^2 $,

$P(\theta_1 < \phi_1 < \theta_i < \phi_i) = \int_{\phi_1}^{2\pi} \int_{\theta_i}^{2\pi} \left( \frac{1}{2\pi} \right)^2 d\phi_i d\theta_i 
= \frac{1}{2} \left(1- \frac{\phi_1}{2\pi} \right)^2 $,

$P(\phi_i < \theta_1 < \phi_1 < \theta_i) = \int_{\phi_1}^{2\pi} \int_{0}^{\theta_1} \left( \frac{1}{2\pi} \right)^2 d\phi_i d\theta_i 
= \frac{\theta_1}{2\pi} \left(1-\frac{\phi_1}{2\pi} \right)$,

\noindent and 
$P(\phi_1 < \theta_i < \phi_i < \theta_1) = \int_{\phi_1}^{\theta_1} \int_{\theta_i}^{\theta_1} \left( \frac{1}{2\pi} \right)^2 d\phi_i d\theta_i 
= \frac{1}{2} \left(\frac{\theta_1-\phi_1}{2\pi} \right)^2$.\\
Therefore,
$$P(X_i = 0) = 
\begin{cases}
    \frac{1}{2 (2\pi)^2}(\theta_1 - \phi_1 + 2\pi)^2 & \text{ for } \theta_1 \leq \phi_1\\
    \frac{1}{2(2\pi)^2}(\theta_1 - \phi_1)^2 & \text{ for } \theta_1 > \phi_1
\end{cases},$$
giving us the required result.
\end{proof}

\begin{remark}
    \label{rmk1}
    If $\Theta,\Phi$ follows any circular distribution with pdf $f(\cdot)$ and cdf $F(\cdot)$, then we have 
    $$P(X_i = 0) = 
    \begin{cases}
    \frac{1}{2} \left( F(\theta_1) - F(\phi_1) \right)^2, & \text{if } \theta_1 > \phi_1\\
    \frac{1}{2} \left( F(\theta_1) - F(\phi_1) + 1 \right)^2, & \text{if } \theta_1 \leq \phi_1.
    \end{cases}$$
\end{remark} 

Using Proposition \ref{fixed_arc_prob} and Remark \ref{rmk1}, we derive the asymptotic distribution of the vertex degree of an RCAG, as stated in Theorem \ref{degree_thm}. 

\begin{theorem}
    \label{degree_thm}
    Let $G \in \mathcal{G}_n$ and $v \in V(G)$ with degree $d(v)$. Then for a fixed $x \in [0,1]$, 
    $$\lim_{n \rightarrow \infty} P(d(v) \leq xn) = 
    \frac{5}{2} - 2\sqrt{2(1-x)} - x -\frac{1}{2}(1-\sqrt{2(1-x)})^2, $$
    if $x > 1/2$, and 
    $$\lim_{n \rightarrow \infty} P(d(v) \leq xn) = 0,$$
    if $x \leq 1/2$.
\end{theorem}

\begin{proof}
    Let $A_1 = [\theta_1,\phi_1]$ be a fixed arc of RCAG. Let $X_i = 1$ if $A_i \cap A_1 \neq \emptyset$ for $ 2 \leq i \leq n$, and $X_i = 0$ otherwise. Then $X_i, X_j$ are independent for $i \neq j$. Let $X = \sum_{i \geq 2} X_i$ be the degree of the vertex corresponding to arc $A_1$. Here, $X_i \sim Bernoulli(p)$ where $p = 1-P(X_i=0)$, with $P(X_i=0)$ obtained from Proposition \ref{fixed_arc_prob}. 
    Then $X|\theta_1,\phi_1 \sim Bin(n-1,p)$ with $E(X) = (n-1)p$ and $Var(X) = (n-1)p(1-p)$. 
    Therefore, $P(X \leq xn | \theta_1,\phi_1) = P \left(Z \leq \frac{n(x-p)+p}{\sqrt{(n-1)p(1-p)}} \right)$ where $Z = \frac{X - (n-1)p}{\sqrt{(n-1)p(1-p)}}$. Rearranging the terms, we get $ P(X \leq xn | \theta_1,\phi_1) = P(Z \leq w)$ where
    $$ w =  \sqrt{n-1} \frac{x-p}{\sqrt{p(1-p)}} + \frac{1}{\sqrt{n-1}} \frac{x-p}{\sqrt{p(1-p)}} + \frac{1}{\sqrt{n-1}} \frac{p}{\sqrt{p(1-p)}} $$ 
    This means, as $n \rightarrow \infty$, $w \rightarrow \infty$ if $x-p > 0$ and $w \rightarrow -\infty$ if $x-p < 0$. Then, by the Central Limit Theorem, as $n \rightarrow \infty$,  $P(X \leq xn | \theta_1,\phi_1) \rightarrow 1$ if $x-p > 0$ and $P(X \leq xn | \theta_1,\phi_1) \rightarrow 0$ if $x-p < 0$. Therefore, 
    $\lim_{n \rightarrow \infty} P(X \leq xn | \theta_1,\phi_1) = I_{x-p > 0} $, where $I_S$ denotes the indicator function of the event $S$.
    Now, $$P(X \leq xn)  = E(I_{X\leq xn}) = E_{\theta_1,\phi_1}[E(I_{X\leq xn}|\theta_1,\phi_1)]$$
    Therefore, by the Bounded Convergence Theorem,
    \begin{align*}
        \lim_{n \rightarrow \infty} P(X \leq xn)  & = E_{\theta_1,\phi_1} \left[ \lim_{n \rightarrow \infty}E(I_{X\leq xn}|\theta_1,\phi_1) \right] \\
        & = E_{\theta_1,\phi_1} (I_{x-p>0}) = P(x-p > 0)
    \end{align*}
    Thus, by putting the value of $p$ from Proposition \ref{fixed_arc_prob},
    \begin{align*}
        \lim_{n \rightarrow \infty} P(X \leq xn) &= P \left(x-1 + \frac{1}{2(2\pi)^2} \left( (\theta_1 - \phi_1) \mod 2\pi \right)^2  > 0\right)\\
        &= P\left( ((\theta_1 - \phi_1) \mod 2\pi) > 2\pi \sqrt{2(1-x)}\right), 
    \end{align*}
for $x \geq \frac{1}{2}$ (as $\theta_1,\phi_1 \leq 2\pi$ 
i.e., $\sqrt{2(1-x)} \leq 1$) and 0 otherwise. Then, for $x\geq \frac{1}{2}$, we can divide it into the following two cases.\\

    \noindent \textbf{Case I:} When $\theta_1 \geq \phi_1$,
    \begin{align*}
        P & \left( ((\theta_1 - \phi_1) \mod 2\pi) > 2\pi \sqrt{2(1-x)} \right) 
         = P\left( \phi_1 <  \theta_1 - 2\pi \sqrt{2(1-x)}\right)\\
        & \hspace{1cm} = \frac{1}{(2\pi)^2} \int_{0}^{2\pi} \int_{0}^{\left(\theta_1 - 2\pi\sqrt{2(1-x)} \right) \vee  0 }  d\phi_1 d\theta_1\\
        & \hspace{1cm} = \frac{1}{(2\pi)^2} \int_{0}^{2\pi} \left[ \left(\theta_1 - 2\pi\sqrt{2(1-x)}\right) \vee 0  \right] d\theta_1
    \end{align*}
where $a \vee b$ denotes $\max\{a,b\}$. Now, substituting $a \vee b = \frac{1}{2} (a+b + |a-b|)$, we get this probability to be
\begin{align*}
    & = \frac{1}{2(2\pi)^2} \int_{0}^{2\pi} \left[ \theta_1 - 2\pi\sqrt{2(1-x)} + |\theta_1 - 2\pi\sqrt{2(1-x)}| \right] d\theta_1 \\
    & = \frac{1}{2(2\pi)^2} \bigl\{ \int_{0}^{2\pi} \left[ \theta_1 - 2\pi\sqrt{2(1-x)} \right] d\theta_1 + \int_{0}^{2\pi\sqrt{2(1-x)}} \left[ 2\pi\sqrt{2(1-x)} - \theta_1 \right] d\theta_1 \\
    & \hspace{6cm} + \int_{2\pi\sqrt{2(1-x)}}^{2\pi} \left[ \theta_1 - 2\pi\sqrt{2(1-x)} \right] d\theta_1 \bigr\}
\end{align*}

\noindent Solving the integral gives
\begin{equation}
    P\left( ((\theta_1 - \phi_1) \mod 2\pi) > 2\pi \sqrt{2(1-x)}\right) = \frac{3}{2} - \sqrt{2(1-x)} - x \hspace{0.5cm} \text{ if } x \geq \frac{1}{2}, \label{eq1}
\end{equation}
and 0 if $x \leq \frac{1}{2}$.\\
    
\noindent\textbf{Case II:} When $\theta_1 < \phi_1$,
\begin{align*}
     P &\left( \left((\theta_1 - \phi_1) \mod 2\pi \right)  >  2\pi \sqrt{2(1-x)} \right) 
    = P\left( \theta_1 - \phi_1 + 2\pi >  2\pi \sqrt{2(1-x)}\right)\\
    & \hspace{1cm} = P\left( \theta_1 > \phi_1 + 2\pi\left( \sqrt{2(1-x)} - 1\right) \right) \\
    & \hspace{1cm} = \frac{1}{(2\pi)^2} \int_{0}^{2\pi} \int_{\left( \phi_1 + 2\pi\left( \sqrt{2(1-x)} - 1\right) \right) \vee 0}^{2\pi } d\theta_1 d\phi_1\\
    & \hspace{1cm} = \frac{1}{(2\pi)^2} \int_{0}^{2\pi} \left\lbrace  2\pi - \left[ \left(\phi_1 + 2\pi\left( \sqrt{2(1-x)} - 1\right) \right) \vee 0 \right] \right\rbrace  d\phi_1\\
    & \hspace{1cm} = 1- \sqrt{2(1-x)} - \frac{1}{2}\left( 1-\sqrt{2(1-x)} \right)^2 \hspace{0.7cm} \text{ if } x \geq \frac{1}{2}, \numberthis \label{eq2}
\end{align*}
and 0 if $x \leq \frac{1}{2}$.

Thus, adding Equations \ref{eq1} and \ref{eq2} gives us the required result.
\end{proof}
  
Theorem \ref{degree_thm} also tells us that for a large value of $n$, the minimum vertex degree of $ G \in \mathcal{G}_n$ is approximately $\frac{n}{2}$.

\begin{corollary}
\label{cor1}
    The asymptotic distribution of the degree of a vertex in an RCAG remains the same irrespective of the choice of distribution of $\Theta_i, \Phi_i$.
\end{corollary}

\begin{proof}
    By proof of Theorem \ref{degree_thm}, we know $\lim_{n \rightarrow \infty} P(X \leq xn)  = P(x-p > 0)$. Then by Remark \ref{rmk1}, 
$$ \lim_{n \rightarrow \infty} P(X \leq xn)  =
\begin{cases}
    P \left(x-1 + \frac{1}{2} \left( F(\theta_1) - F(\phi_1) \right)^2  > 0\right) & \text{if } \theta_1 \geq \phi_1\\
    P \left(x-1 + \frac{1}{2} \left( F(\theta_1) - F(\phi_1) + 1 \right)^2  > 0\right) & \text{if } \theta_1 <  \phi_1
\end{cases}$$
\textbf{Case I:} If $\theta_1 \geq \phi_1$,
\begin{align*}
     P \left(x-1 + \frac{1}{2} \left( F(\theta_1) - F(\phi_1) \right)^2  > 0\right)  &= P \left( F(\phi_1) < F(\theta_1) - \sqrt{2(1-x)} \right)\\
    & = \int_{0}^{1} \int_{0}^{\left( F(\theta_1) - \sqrt{2(1-x)} \right) \vee 0} dF(\phi_1) dF(\theta_1) \\
    & = \int_{0}^{1} \left[ \left( F(\theta_1) - \sqrt{2(1-x)} \right) \vee 0 \right] dF(\theta_1) \\
    & =  \frac{3}{2} - \sqrt{2(1-x)} - x \hspace{1cm} \text{ if } x \geq \frac{1}{2}
\end{align*}
\textbf{Case II:} If $\theta_1 > \phi_1$,
\begin{align*}
    P \left(x-1 + \frac{1}{2} \left( F(\theta_1) - F(\phi_1) + 1 \right)^2  > 0\right)  &= P\left( F(\theta_1) > F(\phi_1) - 1 + \sqrt{2(1-x)}\right) \\
    & \hspace{-1.2cm} = \int_{0}^{1} \int_{ \left( F(\phi_1) - 1 + \sqrt{2(1-x)} \right) \vee 0 }^{1} dF(\theta_1) dF(\phi_1) \\
    & \hspace{-1.2cm} = \int_{0}^{1} \left[ 1- \left[ \left( F(\phi_1) - 1 + \sqrt{2(1-x)} \right) \vee 0 \right]\right] dF(\phi_1) \\
    & \hspace{-1.2cm} = 1- \sqrt{2(1-x)} - \frac{1}{2}\left( 1-\sqrt{2(1-x)} \right)^2 \hspace{0.5cm} \text{ if } x \geq \frac{1}{2}.
\end{align*}
Note that both these probabilities are the same as the ones obtained in Theorem \ref{degree_thm}, giving us the required result.
\end{proof}

This result demonstrates that the vertex degree distribution in an RCAG depends only on the number of vertices $n$, and not on the underlying distribution of the observations. We now proceed to derive the joint distribution of vertex degrees in an RCAG. These results will later be instrumental in Section \ref{DD_algo}, where we develop our test for assessing randomness.

\begin{theorem}
    \label{2_arc_fix}
    Let $d_1$ and $d_2$ be the degrees of any two vertices of $G \in \mathcal{G}_n$. Then,
    $$\lim_{n \rightarrow \infty}P(d_1 \leq xn,d_2 \leq xn) =  \left[ \lim_{n \rightarrow \infty}P(d_j \leq xn) \right]^2 =  \lim_{n \rightarrow \infty}P(d_1 \leq xn) \lim_{n \rightarrow \infty}P(d_2 \leq xn) ,$$
for $j=1,2$. 
\end{theorem}

\begin{proof}
    Let $A_1 = [\theta_1,\phi_1]$ and $A_2 = [\theta_2,\phi_2]$ be two fixed arcs of $G \in \mathcal{G}_n$. Let $X_i^{(1)} = 1$ if $A_i \cap A_1 \neq \emptyset$ for $ 3 \leq i \leq n$, and $X_i^{(1)} = 0$ otherwise and $X_i^{(2)} = 1$ if $A_i \cap A_2 \neq \emptyset$ for $ 3 \leq i \leq n$, and $X_i^{(2)} = 0$ otherwise. Then $X_i^{(1)}, X_j^{(1)}$ are independent for $i \neq j$ and $X_i^{(2)}, X_j^{(2)}$ are independent for $i \neq j$. Let $X_{12} =1$ if $A_1 \cap A_2 \neq \emptyset$  and $X_{12} =0$ otherwise.  
Let $d_1 = \sum_{i \geq 3} X_i^{(1)} + X_{12}$ and $d_2 = \sum_{i \geq 3} X_i^{(2)} + X_{12}$ be the degree of the vertex corresponding to arcs $A_1$ and $A_2$ respectively. Here $X_i^{(j)} \sim Bernoulli(p_j)$ for $3 \leq i \leq n$, $j=1,2$ where $p_j = P(X_i^{(j)}=1|\theta_j,\phi_j)$ can be obtained from Proposition \ref{fixed_arc_prob}. 
Then $d_j \sim Bin(n-2,p_j)$ if $X_{12}=0$ and $d_j \sim Bin(n-2,p_j)+1$ if $X_{12}=1$. 

If $X_{12}=0$, then we can write $P(d_1 \leq xn,d_2 \leq xn|\theta_1,\phi_1,\theta_2,\phi_2) = P(Z_1\leq w_1, Z_2 \leq w_2)$ where $Z_j = \frac{d_j - (n-2)p_j}{\sqrt{(n-2)p_j(1-p_j)}}$ and 
$$ w_j =  \sqrt{n-2} \frac{x-p_j}{\sqrt{p_j(1-p_j)}} + \frac{2}{\sqrt{n-2}} \frac{x-p_j}{\sqrt{p_j(1-p_j)}} + \frac{2}{\sqrt{n-2}} \frac{p_j}{\sqrt{p_j(1-p_j)}}.$$ 
Else, if $X_{12}=1$, then we can write $P(d_1 \leq xn,d_2 \leq xn|\theta_1,\phi_1,\theta_2,\phi_2) = P(Z_1\leq w_1, Z_2 \leq w_2)$ where $Z_j = \frac{d_j - (n-2)p_j-1}{\sqrt{(n-2)p_j(1-p_j)}}$ and 

$$ w_j =  \sqrt{n-2} \frac{x-p_j}{\sqrt{p_j(1-p_j)}} + \frac{2}{\sqrt{n-2}} \frac{x-p_j}{\sqrt{p_j(1-p_j)}} + \frac{1}{\sqrt{n-2}} \frac{2p_j-1}{\sqrt{p_j(1-p_j)}}.$$ 
This means, as $n \rightarrow \infty$, $w_j \rightarrow \infty$ if $x-p_j > 0$ and $w \rightarrow -\infty$ if $x-p_j < 0$ for $j=1,2$ irrespective of whether $X_{12}=0$ or $X_{12}=1$. Then, by the Central Limit Theorem, as $n \rightarrow \infty$,  $P(d_1 \leq xn,d_2 \leq xn | \theta_1,\phi_1,\theta_2,\phi_2) \rightarrow 1$ if $x-p_1 > 0$ and $x-p_2 > 0$ and $P(d_1 \leq xn,d_2 \leq xn | \theta_1,\phi_1) \rightarrow 0$ if $x-p_1 < 0$ or $x-p_2 < 0$. Therefore,
$\lim_{n \rightarrow \infty} P(d_1 \leq xn,d_2 \leq xn | \theta_1,\phi_1,\theta_2,\phi_2) = I_{(x-p_1 > 0) \: \cap \: (x-p_2>0)}.$ Proceeding in a similar manner as in Theorem \ref{degree_thm}, we get $ \lim_{n \rightarrow \infty} P(d_1 \leq xn,d_2 \leq xn ) = P\left( x-p_1 > 0 \text{ and } x-p_2>0 \right)$. This equals 
$P\left( ((\theta_1 - \phi_1) \mod 2\pi) > 2\pi \sqrt{2(1-x)} \text{ and } ((\theta_2 - \phi_2) \mod 2\pi) > 2\pi \sqrt{2(1-x)} \right)$ by \\ 
using Proposition \ref{fixed_arc_prob}. 
Since both of these events do not depend on each other, we get $\lim_{n \rightarrow \infty} P(d_1 \leq xn,d_2 \leq xn )$ $= P\left( ((\theta_1 - \phi_1) \mod 2\pi) > 2\pi \sqrt{2(1-x)}  \right) $  $\times$ \\
$P \left( ((\theta_2 - \phi_2) \mod 2\pi) > 2\pi \sqrt{2(1-x)} \right)$ $=P(d_1 \leq xn) P(d_2\leq xn)$.
\end{proof}


\subsection{Connectedness}
\label{max_degree}


We now investigate the number of connected components in an RCAG. Let $|A|$ denote the length of an arc $A$.

\begin{lemma}
\label{arc_len}
    Let $A = [\Theta,\Phi]$ be an arc of an RCAG with $\Theta,\Phi$ i.i.d. with  pdf $f(\cdot)$ and cdf $F(\cdot)$. Then 
    $$P(|A| \leq x) = \frac{1}{2} - F(2\pi-x) + \int_{0}^{2\pi-x} F(\theta+x) f(\theta) d\theta + \int_{2\pi-x}^{2\pi} F(\theta+x-2\pi) f(\theta) d\theta.$$
\end{lemma}

\begin{proof}
    Let $A=[\theta,\phi]$ be an arc. Then, for a fixed value of $\theta$, $|A| \leq x$ if $\phi \leq \theta + x$ (as in a unit circle angle subtended by $A$  is equal to the length of $A$, i.e.,  $|A|$). Therefore, for $\theta \leq \phi$,
\begin{align*}
    P(|A| \leq x) &= \int_{0}^{2\pi-x} \int_{\theta}^{\theta+x} f(\theta) f(\phi) \: d\phi \: d\theta \\
    & = \int_{0}^{2\pi-x} F(\theta+x) f(\theta) \: d\theta - \frac{F^2(2\pi-x)}{2}
\end{align*}
and for $\theta > \phi$,
\begin{align*}
    P(|A| \leq x) &= \int_{2\pi-x}^{2\pi} \int_{\theta}^{(\theta+x) \text{ mod } 2\pi} f(\theta) f(\phi) \: d\phi \: d\theta \\
    & = \int_{2\pi-x}^{2\pi} \int_{\theta}^{2\pi} f(\theta) f(\phi) \: d\phi \: d\theta + \int_{2\pi-x}^{2\pi} \int_{0}^{\theta + x -2\pi} f(\theta) f(\phi) \: d\phi \: d\theta \\
    & = \frac{1}{2} - F(2\pi-x) + \frac{F^2(2\pi-x)}{2} + \int_{2\pi-x}^{2\pi} F(\theta + x -2\pi) f(\theta) \: d\theta.
\end{align*}
Adding both gives the required result.
\end{proof}

\begin{remark}
\label{arc_len_cunif}
    If $\Theta,\Phi \sim$ Circular Uniform distribution then $P(|A| \leq x) = \frac{x}{2\pi}$ for $0  \leq x \leq 2\pi$.
    
\end{remark}

\begin{proposition}
    \label{max_len}
    Let the observations $\Theta,\Phi \sim$ pdf $f(\cdot)$ and cdf $F(\cdot)$ with $f$ bounded on the unit circle. Let $a=\alpha n$ where $0< \alpha <<1$. Then, for almost all $G \in \mathcal{G}_n$, there exists an arc $A$ such that $|A| > 2\pi - \frac{a}{n}$, for large values of $n$. 
\end{proposition}

\begin{proof}
    Let $A_i$ be the arcs corresponding to the vertices of $G$ for $1 \leq i \leq n$. Here we want to find $\lim_{n \rightarrow \infty} P \left( \text{at least one } |A_i| > 2\pi - \frac{a}{n} \right)$.
By Lemma \ref{arc_len}, we have
$$P \left(|A_i| \leq 2\pi- \frac{a}{n} \right) =  \frac{1}{2} - F \left(\frac{a}{n} \right) + \int_{0}^{\frac{a}{n}} F \left( \theta + 2\pi- \frac{a}{n} \right) f(\theta) d\theta + 
\int_{\frac{a}{n}}^{2\pi} F \left( \theta - \frac{a}{n} \right) f(\theta) d\theta.$$
This can be rearranged as
$$P \left(|A_i| \leq 2\pi- \alpha \right) = 1- \left( \frac{1}{2} + F (\alpha) -I_1 - I_2 \right),$$
where 
$I_1 = \int_{0}^{\alpha} F ( \theta + 2\pi- \alpha) f(\theta) d\theta$ and $I_2 = \int_{\alpha}^{2\pi} F (\theta -\alpha ) f(\theta) d\theta$.
Therefore, 
$$P \left( \text{all } |A_i| \leq 2\pi- \alpha \right) = \left[ 1- \left( \frac{1}{2} + F (\alpha) -I_1 - I_2 \right) \right]^n.$$

For small values of $ \alpha$, we can approximate $I_1$ in a small neighbourhood around 0 as  $F(2\pi - \alpha)f(0)\alpha \approx [F(2\pi) - \alpha f(2\pi)] f(0)\alpha$ by the Taylor series approximation. Since $f(2\pi) = f(0)$, this gives $I_1 \approx \alpha f(0) - \alpha^2 f^2(0)$. 
Similarly, $I_2  \approx  E[F(\Theta-\alpha)] \approx E[F(\Theta)-\alpha f(\Theta)] \approx 1/2 - \alpha E[f(\Theta)]$. 
Therefore, as $F(\alpha) \approx \alpha f(0)$, $ 1/2 + F (\alpha) -I_1 - I_2  \approx \alpha^2 f^2(0) + \alpha E[f(\Theta)]$. 
Now, for any $\epsilon >0,$ and $ a = \alpha n, \alpha>0$, let $N_1  > \{a^2 f^2(0) + aE[f(\Theta)]\}/\epsilon \in \mathbb{N}$. Then, for all $n > N_1$, $|1/2 + F (\alpha) -I_1 - I_2| < \epsilon$. Therefore, $\lim_{n \rightarrow \infty} \left( 1/2 + F (\alpha) -I_1 - I_2 \right) = 0$.
Also, we know that $\lim_{x\rightarrow a} (1+f(x))^{g(x)} = e^{\lim_{x\rightarrow a} f(x)g(x)} $ if $\lim_{x\rightarrow a}f(x) =0 $ and $\lim_{x\rightarrow a}g(x) = \infty $.
Thus, 
\begin{align*}
    \lim_{n \rightarrow \infty} P \left( \text{all } |A_i| \leq 2\pi- \alpha \right) &= e^{\lim_{n \rightarrow \infty} - \left( \frac{1}{2} + F (\alpha) -I_1 - I_2 \right) \times n} \approx e^{\lim_{n \rightarrow \infty} - \left( \alpha^2 f^2(0) + \alpha E[f(\Theta)] \right) \times n}
\end{align*}

Now, for any $M >0,\alpha>0$ small, let $N_2 > M/\{\alpha^2 f^2(0) + a E[f(\Theta)]\} \in \mathbb{N}$. Then, for all $n>N_2$, $\left\{ (\alpha^2 f^2(0) + a E[f(\Theta)]) \times n \right\} > M$. Therefore, $\lim_{n \rightarrow \infty} \left\{ \left( \alpha^2 f^2(0) + \alpha E[f(\Theta)] \right) \times n \right\} $ $ = \infty$. Thus, 
$ \lim_{n \rightarrow \infty} P \left( \text{all } |A_i| \leq 2\pi- \alpha \right) = e^{-\infty} = 0$, giving us the required result.
\end{proof}

Using Lemma \ref{arc_len} and Proposition \ref{max_len}, we derive Theorem  \ref{max_deg_lem} given below, which tells us the maximum vertex degree of an RCAG when the number of vertices is large.

\begin{theorem}
    \label{max_deg_lem}
     Let $G \in \mathcal{G}_n$. Then for large values of $n$, the maximum vertex degree of $G$ is $n-1$ with probability 1.
\end{theorem}

\begin{proof}
    Let $A_i$'s be the arcs for $G \in \mathcal{G}_n$, for $1 \leq i \leq n$. Then, by Proposition \ref{max_len}, we know that there exists an arc in $G$ of length greater than $2\pi - \frac{a}{n}$ with probability one for large values of $n$. 
Let us fix this arc as $A_1 = (\theta_1,\phi_1)$. 
Let $X_i = 1$ if $A_i \cap A_1 \neq \emptyset$ for $ 2 \leq i \leq n$, and 0 otherwise. Then by Remark \ref{rmk1},
$$P(X_i = 0) = 
\begin{cases}
\frac{1}{2} \left( F(\theta_1) - F(\phi_1) \right)^2, & \text{if } \theta_1 > \phi_1\\
\frac{1}{2} \left( F(\theta_1) - F(\phi_1) + 1 \right)^2, & \text{if } \theta_1 \leq \phi_1.
\end{cases}$$
Now, for $\theta_1 > \phi_1$, $|A_1| =2\pi - (\theta_1 - \phi_1) > 2\pi - \frac{a}{n}$, which gives $\theta_1 < \phi_1 + \frac{a}{n}$. Using this with the Taylor series approximation for large values of $n$, we get
\begin{align*}
    P(X_i =0) &< \frac{1}{2} \left( F \left(\phi_1 + \frac{a}{n} \right) - F(\phi_1) \right)^2 \\
    & \approx \frac{1}{2} \left( F (\phi_1) + \frac{a}{n}f(\phi_1) - F(\phi_1) \right)^2 \\
    &= \frac{1}{2} \left( \frac{a}{n}f(\phi_1) \right)^2 \rightarrow 0  \text{ as } n \rightarrow \infty.
\end{align*}

Similarly, for $\theta_1 \leq \phi_1$, $|A_1| =\phi_1 -\theta_1 > 2\pi - \frac{a}{n}$ which gives $ \phi_1 \geq \theta_1+ 2\pi - \frac{a}{n}$. Since $\phi_1 \leq 2\pi$, we will have $\theta_1 \leq \frac{a}{n}$, meaning $\theta_1$ could only take very small values. So, $F(\theta_1) \approx \theta_1 f(0)$. Again, using these with Taylor series approximation, we get
\begin{align*}
    P(X_i =0) &< \frac{1}{2} \left( F(\theta_1) -F\left(\theta_1 + 2\pi - \frac{a}{n} \right) + 1 \right)^2 \\
    & \approx \frac{1}{2} \left( \theta_1 f(0) - 1 + \left(\frac{a}{n} - \theta_1 \right) f(2\pi)   + 1 \right)^2 \\
    &= \frac{1}{2} \left( \frac{a}{n}f(0) \right)^2 \rightarrow 0  \text{ as } n \rightarrow \infty.
\end{align*}

Therefore, $\lim_{n\rightarrow \infty}P(X_i=0) = 0$ for all $2 \leq i \leq n$. Let the degree of $A_1$ be $X = \sum_{i=2}^{n-1} X_i$. Then $\lim_{n \rightarrow \infty} P(X = n-1)  = \lim_{n \rightarrow \infty}P(X_i = 1, \: \forall \: 2 \leq i \leq n) = $ \newline $\lim_{n \rightarrow \infty} \prod_{i=2}^{n-1}P (X_i = 1) = \lim_{n \rightarrow \infty} (P (X_2 = 1))^{n-1}$, as $X_i$ are i.i.d. Bernoulli($P(X_2 = 1)$). Then for $\theta_1 > \phi_1$,
\begin{align*}
    \lim_{n \rightarrow \infty} P(X = n-1) &= \lim_{n \rightarrow \infty} \left( 1-\frac{a^2}{n^2}f^2(\phi_1) \right)^{n-1}\\
    & = e^{\lim_{n \rightarrow \infty}  -(n-1)\frac{a^2}{n^2}f^2(\phi_1) } = e^0 = 1.
\end{align*}
Similarly, for $\theta_1 \leq \phi_1$,
\begin{align*}
    \lim_{n \rightarrow \infty} P(X = n-1) &= \lim_{n \rightarrow \infty} \left( 1-\frac{a^2}{n^2}f^2(0) \right)^{n-1}\\
    & = e^{\lim_{n \rightarrow \infty}  -(n-1)\frac{a^2}{n^2}f^2(0)} = e^0 = 1.
\end{align*}
Hence the result.
\end{proof}

Theorem \ref{max_deg_lem} establishes that for sufficiently large 
$n$, almost all $G \in \mathcal{G}_n$ consists of a single component, implying that it is a connected graph with a probability approaching 1.

\subsection{Hamiltonian}


Once we have obtained the result for the degree of a vertex in RCAG, we use its relation with the Hamiltonicity property of a graph to identify the presence of a Hamiltonian cycle in an RCAG. The exact result is as follows.

\begin{theorem}
    \label{ham}
    Let $G \in \mathcal{G}_n$. Then $G$ is Hamiltonian with probability 1 for large $n$.
\end{theorem}

\begin{proof}
    Theorem \ref{degree_thm} tells us that $\lim_{n \rightarrow \infty} P(d(v) \leq \frac{n}{2}) = 0$ for all vertices $v$ in $G$. This means $\lim_{n \rightarrow \infty} P(\delta(G) \geq \frac{n}{2}) = 1$. Then the result follows from the classic result by \cite{dirac}, which states that for a simple graph $G$ with $n \geq 3$, if the minimum degree, $\delta(G) \geq \frac{n}{2}$, then $G$ contains a Hamiltonian cycle.
\end{proof}

The next natural question is how many Hamiltonian cycles does an RCAG contain? \cite{Hamiltonian_number} show that $n$-vertex Dirac graphs contain at least $c^n n!$ distinct Hamiltonian cycles for some small positive constant $c$. \cite{ham_dirac} shows that this $c \approx \delta(G)/n$ where $\delta(G)$ denotes the minimum degree of a vertex of graph $G$. 
This gives us our result for RCAGs as follows.

\begin{lemma}
    \label{no_ham}
    For large values of $n$, $G \in \mathcal{G}_n$ contains at least $\frac{n!}{(2 + o(1))^n}$ Hamiltonian cycles almost surely.   
\end{lemma}

\begin{proof}
    \cite{ham_dirac} shows that for an $n$-vertex Dirac graph, there exists at least $\frac{n!}{(2+o(1))^n}$ Hamiltonian cycles. Since the RCAG is an $n$-vertex Dirac graph, the result directly follows.
\end{proof}

Detecting a Hamiltonian cycle in a graph is an NP-hard problem \citep{np_hard}, making its use for randomness testing computationally expensive. Consequently, such an approach is currently impractical. However, if efficient algorithms for Hamiltonian cycle detection are developed in the future, they could serve as the basis for a new randomness test.

\section{Tests for Randomness}
\label{ToR}


In Section \ref{RCAG}, we saw that the characteristics like the probability of an edge between any two vertices, the maximum degree of a vertex, the vertex degree distribution, and the Hamiltonian property of an RCAG are invariant with respect to the choice of the distribution of observations, and these properties depend solely on the fact that the observations are randomly generated (i.e. mutually independent). With this understanding, we devise two tests for checking the randomness of a given circular dataset. The first test uses the probability of an edge between any two vertices, while the second one is based on the vertex degree distribution.


\subsection{Edge Probability Test}
\label{EP_algo}


Suppose we are given a set of $m$ observations on the circle, represented as $\theta_1, \theta_2, \ldots, \theta_{m-1},$ $ \theta_{m}$. Our objective is to build an RCAG for this set of observations so that its properties can be utilized in our analysis.  First, let $m=4n$. Then, we construct $2n$ arcs $A_j$ from these observations, defined as 
$A_j= [\theta_{2j-1}, \theta_{2j}], 1\leq j \leq 2n$. 
These arcs correspond to $2n$ vertices of the RCAG.
Next, we randomly form $n$ pairs of vertices from these $2n$ vertices, ensuring that each vertex is part of only one pair. We associate an edge between the two vertices in a pair if the intervals associated with them have a nonempty intersection (i.e. if $A_i$ and $A_j$ are the two intervals associated with the two vertices in a pair, then there is an edge between these two vertices if and only if $A_i \cap A_j \ne \emptyset$).
Define $Y_i = 1$ if the $i^{th}$ pair of vertices does not have an edge between them and $Y_i = 0$ otherwise. 

From Section \ref{RCAG}, we find that if the observations are generated randomly, the probability that there exists an edge between any two vertices of an RCAG is $5/6$. Thus, $Y_i \sim Bernoulli (1/6)$ for all $i$. Since the pairs are randomly formed and each vertex is part of only one pair, by Lemma \ref{lemma_edge_dept}, the random variables $Y_i, 1 \le i \le n$ are mutually independent. Let $Y = \sum_{i=1}^{n} Y_i$ be the number of pairs of vertices whose associated arcs do not intersect. Then $ Y \sim Bin (n,1/6)$. We now use this fact to construct a test of randomness. 

Let $G$ be an RCAG constructed by following the above procedure. Let $p$ be the probability that an edge between two randomly chosen vertices in $G$ does not exist. If the observations are mutually independent, then  $p=1/6$ as noted above. Thus, to test for randomness, we test  
$$H_0 : p=\frac{1}{6}  \text{ \quad versus \quad}  H_1: p \neq \frac{1}{6}.$$

Let $\hat{p} = 1/n \sum_{i=1}^{n} y_i$ be the proportion of pairs whose vertices are not joined by an edge.  Using $\hat{p}$ as the test statistic, we propose the test: Reject $H_0$ if $\left| \hat{p} - 1/6 \right| > c$, where $c$ is chosen such that the level of the test is $\alpha$.
It is well known that for large values of $n$, $\hat{p}$ follows a Normal$\left(1/6, 5/(36n) \right)$ distribution, using which the value of $c$ can be computed. For small values of $n$, an exact test (possibly randomized) can be constructed (refer to Chapter 9 in \cite{Book_Rohatgi}, pp. 429-432).  
Define the randomized test $\varphi(y_1, \ldots, y_n)$ of size $\alpha$ as 
$\varphi (y_1, \ldots, y_n) = 1$ if $\sum_{i=1}^{n} y_i \in K_1$, $\varphi (y_1, \ldots, y_n) = \gamma$ if $\sum_{i=1}^{n} y_i \in K_2$ and $\varphi (y_1, \ldots, y_n) = 0$ otherwise, where $K_1, K_2 $ and $\gamma$ are chosen such that $E_{H_0}[\varphi] = \alpha$. 
Thus we reject $H_0$ with probability one if $\sum_{i=1}^{n} y_i$ takes values in $K_1$, with probability $\gamma \in (0,1)$ if $\sum_{i=1}^{n} y_i$ takes values in $K_2$ and do not reject $H_0$ otherwise.

When $m$ is not a multiple of four, i.e., $m = 4n+k$ for $k =1,2,3$, the above-mentioned method cannot be used directly. To resolve this issue, one strategy involves omitting $k$ observations to achieve a dataset of length $4n$, enabling the use of the earlier method. If $k=3$, another ad-hoc approach is to select one observation randomly from the $m$ observations and repeat it in the dataset. These ad-hoc approaches have the shortcoming that the results may vary depending on the dropped or repeated observation(s). Instead of the above two ad-hoc approaches, we suggest a method that involves organizing the observations into groups of size $4n$ and repeating the testing process $k+1$ times. For instance, if $k=3$, groups would be formed as follows: Group-1: $\theta_1, \dots, \theta_{4n}$, Group-2: $\theta_2, \dots, \theta_{4n+1}$, Group-3: $\theta_3, \dots, \theta_{4n+2}$ and Group-4: $\theta_4, \dots, \theta_{4n+3}$. We apply the test described in the preceding paragraph on each of these groups separately and 
apply the Benjamini-Hochberg method \citep{BY_correction} of multiple testing to obtain the adjusted p-values. If the least adjusted p-value is greater than the specified level of significance, we do not reject $H_0$. We call this test the \textit{RCAG-Edge Probability (RCAG-EP) Test}. Algorithm \ref{algo_RCAG_EP} describes this RCAG-EP test for randomness. 

\begin{algorithm}
    \caption{RCAG-Edge Probability Test}
    \begin{algorithmic}[1]
        \State  \textbf{Input:} A univariate series of $m$ observations $\theta_1,\theta_2,\dots, \theta_{m-1}, \theta_{m}$.
        \If {$m=4n$}
        \State Build an RCAG with $2n$ vertices where  $A_j = [\theta_{2j-1},\theta_{2j}] , 1\leq j \leq 2n$
	\State Form $n$ pairs of vertices out of these $2n$ vertices at random, such that each vertex is there in only one pair. 
        \State Form an edge between the two vertices in a pair if the arcs associated with them have a nonempty intersection.
        \State For testing $H_0 : p=\frac{1}{6} \text{ vs } H_1: p \neq \frac{1}{6}$ ($p$ is the probability that there does not exist an edge between any two vertices in an RCAG),
        compute $\hat{p}$, the proportion of pairs whose vertices are not joined by an edge. 
        \If {$n$ is large} \hspace{0.1cm}
         Reject $H_0$ if $|\hat{p} - \frac{1}{6}| > c$, where $c = \sqrt{\frac{5}{36n}}  z_{\frac{\alpha}{2}}$ for level $\alpha$.
        \Else \hspace{0.1cm}
        Reject $H_0$ with probability 1 if $n \hat{p} \in K_1$, with probability $\gamma$ if $n \hat{p} \in K_2$ and do not reject $H_0$ otherwise, where $K_1,K_2,\gamma$ are determined based on the level $\alpha$ 
        \EndIf
        \ElsIf {$m = 4n+k, k =1,2,3$}
        \State Group observations into $k+1$ groups $G_1, \ldots, G_{k+1}$ of size $4n$ each, where $G_1=\{\theta_1, \ldots, \theta_{4n}\}$, $G_2= \{\theta_2, \ldots, \theta_{4n+1}\}$, \dots, $G_{k+1} = \{\theta_{k+1}, \ldots, \theta_{4n+k}\}$
   
        \State Apply Steps 2-9 to each of the groups $G_i$ to obtain $k+1$ p-values.
        \State Apply the Benjamini and Hochberg procedure to obtain the adjusted p-values.
        \State If the least adjusted p-value is $> \alpha$, do not Reject $H_0$ else Reject $H_0$
        \EndIf
    \end{algorithmic}
    \label{algo_RCAG_EP}
\end{algorithm}


\subsection{Degree Distribution Test}
\label{DD_algo}


Another key characteristic of the RCAG that can be leveraged to test for randomness is the degree distribution of a vertex. In Section \ref{degree}, we established that the degree distribution of a vertex in an RCAG remains the same regardless of the underlying distribution of observations. Consequently, for a fixed number of vertices, the theoretical degree distribution can be directly derived from Theorem \ref{degree_thm}. This implies that if we construct an RCAG for a given set of observations and the observations are mutually independent, the resulting empirical degree distribution should align closely with the theoretical distribution from Theorem \ref{degree_thm}. Therefore, the similarity between these distributions can serve as an indicator of randomness in the observations.

\begin{figure}[t]
\centering 
    \begin{subfigure}{0.33\textwidth}
        \centering\includegraphics[scale=0.35]{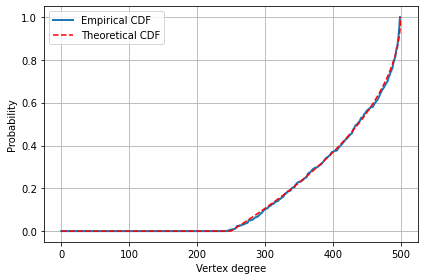}
        \caption{CDF Circular Uniform}
    \end{subfigure}\hfil 
    \begin{subfigure}{0.33\textwidth}
        \centering\includegraphics[scale=0.35]{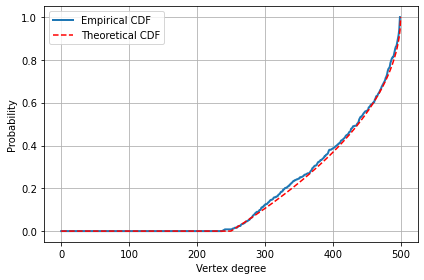}
        \caption{CDF von Mises}
    \end{subfigure}\hfil 
        \begin{subfigure}{0.33\textwidth}
        \centering\includegraphics[scale=0.35]{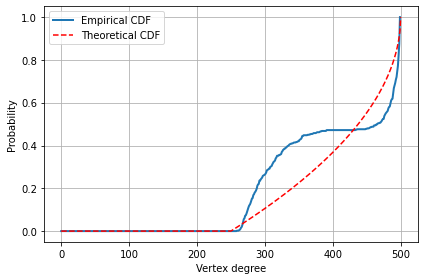}
        \caption{CDF LAR(1) series}
    \end{subfigure}\hfil 
    \begin{subfigure}{0.33\textwidth}
        \centering\includegraphics[scale=0.35]{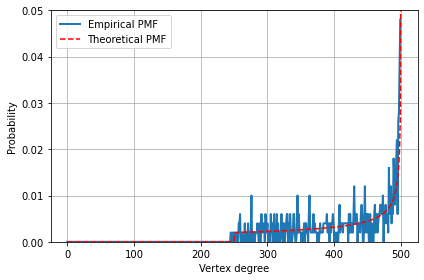}
        \caption{PMF Circular Uniform}
    \end{subfigure}\hfil 
    \begin{subfigure}{0.33\textwidth}
        \centering\includegraphics[scale=0.35]{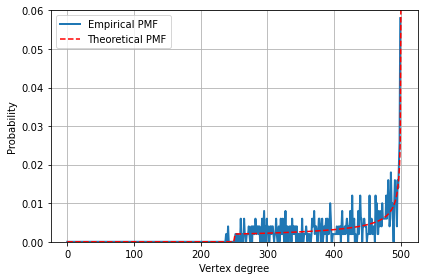}
        \caption{PMF von Mises}
    \end{subfigure}\hfil 
    \begin{subfigure}{0.33\textwidth}
        \centering\includegraphics[scale=0.35]{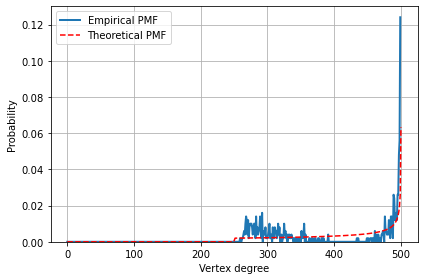}
        \caption{PMF LAR(1) series}
    \end{subfigure}\hfil 
\caption{Comparison between theoretical and empirical degree distributions of RCAGs with 1000 observations when the data is generated randomly from circular uniform distribution in (a) and (d), randomly from von Mises distribution in (b) and (e), and LAR(1) process in (c) and (f).}
\label{fig_dd_exmpl}
\end{figure}

Figure \ref{fig_dd_exmpl} illustrates this concept with 1000 observations generated under three different scenarios: (i) randomly from a Circular Uniform distribution, (ii) randomly from a von Mises distribution, and (iii) from a \textit{Linked Autoregressive} (LAR) process (described in detail in Section \ref{exp}). The figure demonstrates that when the observations are truly random, the empirical and theoretical degree distributions closely match. However, for the LAR(1), where the observations exhibit dependence, a noticeable deviation from the theoretical degree distribution is observed.

Formally, suppose we are given a set of $m$ observations  $\theta_1,\theta_2,\dots,\theta_{m}$ on the unit circle. First, let $m=2n$. Then, as mentioned in Section \ref{EP_algo}, we build an RCAG corresponding to the given observations as follows. The $j^{th}$ vertex of the RCAG $G$ corresponds to the arc $A_j = [\theta_{2j-1},\theta_{2j}]$, $1\leq j \leq n$. Let $\hat{F}_n$ be the empirical degree distribution of this graph $G$, and $F^*$ be the theoretical degree distribution of RCAG with $n$ vertices.
Let $\hat{F_n}(x)  = 1/n\sum_{j=1}^{n} I(d_j\leq xn)$ where $d_j$ is the degree of vertex $j$.
Since the $d_j$'s are not i.i.d., we cannot claim that $\hat{F_n}$ converges to $F^*$ uniformly using the Glivenko-Cantelli theorem. 

By Theorem \ref{degree_thm}, when $\theta_i$'s are mutually independent we have $\lim_{n \rightarrow\infty} P(d_j \leq xn)= F^*(x)$, for every $1 \le j \le n$. Hence, for any given $\epsilon>0, \exists \: N_j$ such that $\forall \: n \geq N_j$, $|P(d_j \leq xn) - F^*(x)| < \epsilon$. Now, define $k(n)$ to be the number of $j$ for which $n \geq N_j$. 

\begin{theorem}
    \label{Fn_convg}
    Let $\hat{F}_n(x)$ be as defined above and $\lim_{n \rightarrow \infty} \frac{k(n)}{n} = 1$. Then $\hat{F}_n(x)$ converges in probability to $F^*(x)$ for every $x\in [0,1]$.
\end{theorem}

\begin{proof}
    Let $t=xn$. Then as $n\rightarrow\infty, t \rightarrow \infty$. We have 
$$E[\hat{F}_n(x)] = \frac{1}{n}\sum_{i=1}^{n} E[ I(d_j\leq xn)] = \frac{1}{n}\sum_{j=1}^{n} P(d_j \leq t)$$

Let $S_1 = \{j \leq n : |P(d_j \leq xn) - F^*(x)|< \epsilon \}$ and $S_2 = \{j \leq n : |P(d_j \leq xn) - F^*(x)| \geq \epsilon \} $. Then $|S_1|= k(n)$ and $|S_2| = n-k(n)$. Now we can write the expression of $E[\hat{F}_n(x)]$ as
\begin{align*}
    E[\hat{F}_n(x)] &= \frac{1}{n}\sum_{j=1}^{n} \left[ P(d_j \leq t) -F^*(x) \right] + F^*(x)\\
    &= \frac{1}{n}\sum_{j\in S_1} \left[ P(d_j \leq t) -F^*(x) \right] +\frac{1}{n}\sum_{j\in S_2} \left[ P(d_j \leq t) -F^*(x) \right] + F^*(x)\\
    & \leq \frac{k(n)}{n} \epsilon + 2 \left( \frac{n-k(n)}{n} \right) + F^*(x).
\end{align*}
Since $\lim_{n \rightarrow \infty} \frac{k(n)}{n}=1$, we get $\lim_{n \rightarrow \infty}E[\hat{F}_n(x)] = F^*(x)$. Now, 
\begin{align*}
    & Var(\hat{F}_n(x)) = \frac{1}{n^2} Var \left( \sum_{j=1}^n I(d_j \leq t)\right) \\
    & = \frac{1}{n^2} \left( \sum_{j=1}^n Var \left(I(d_j \leq t)\right) + 2 \sum_{j=1}^n\sum_{i <j} Cov (I(d_i \leq t)I(d_j \leq t))\right) \\
    &= \frac{1}{n^2} \left( \sum_{j=1}^n P(d_j \leq t) (1-P(d_j \leq t)) + 2 \sum_{j=1}^n\sum_{i <j} \left[ P(d_i \leq t,d_j \leq t) -  P(d_i \leq t) P(d_j \leq t) \right] \right)\\
    &\leq \frac{1}{4n} + \frac{1}{n^2} \left( 2 \sum_{j=1}^n\sum_{i <j} \left[ P(d_i \leq t,d_j \leq t) -  P(d_i \leq t) P(d_j \leq t) \right] \right)
\end{align*}
Then by Theorem \ref{2_arc_fix}, we can conclude that $\lim_{n \rightarrow \infty} Var(\hat{F}_n(x)) =0$. Thus, by using Chebyshev's Inequality, we get the required result.
\end{proof}

From Theorem \ref{Fn_convg}, we can conclude that when the observations are mutually independent, $\hat{F}_n(x)$ converges to $F^*(x)$ in probability for every $x\in [0,1]$.
We now use this property to construct a test for randomness based on the vertex degree distribution.
We base the test on the Hellinger distance $HD$ between the theoretical and empirical pmfs of the vertex degree.The theoretical pmf is $f^*(i) = F^*(i/n)-F^*((i-1)/n)$ and the empirical pmf is $\hat{f}_n(i) = \hat{F}_n(i/n) -\hat{F}_n((i-1)/n)$. Therefore, $$HD = \frac{1}{\sqrt{2}} \sum_{i=1}^n \left( \sqrt{\hat{f}_n(i)} - \sqrt{f^*(i)}\right)^2$$ is our test statistic. We define the rejection rule as: Reject the null hypothesis that the observations are random (mutually independent) if the value of $HD$ is greater than some threshold value $C_\alpha$ chosen based on the level of significance $\alpha$. 
The null distribution of HD seems to be intractable to derive theoretically. Hence, we take recourse to simulation to determine the value of $C_\alpha$. 

When $m$ is not a multiple of two, one way to test for randomness is to randomly remove one observation, reducing the dataset to size 
$2n$ so that we can apply the method described above. However, since the result may depend on the observation removed, we suggest an alternative procedure similar to the one used for the RCAG-EP test. For $m=2n+1$, we create two groups: Group-1: $\theta_1, \dots, \theta_{2n}$ and Group-2: $\theta_2, \dots, \theta_{2n+1}$.  
We compute the test statistics $HD_1$ and $HD_2$ for these two groups separately.  We now reject the null hypothesis of randomness if any of the test statistics is greater than $C_{\alpha/2}$. 

We call this test the \textit{RCAG-Degree Distribution (RCAG-DD) Test}. Algorithm \ref{algo_RCAG_DD} describes the RCAG-DD test for randomness.

In this paper, to calculate the value of $C_\alpha$, we generate 1000 random series of length $2n$ from the Circular Uniform distribution and compute test statistics $HD^{(1)}, HD^{(2)}, \dots, $ $ HD^{(1000)}$ for each of these generated random series. 
Then, for a test at the level of significance $\alpha$, $C_\alpha =100(1-\alpha)^{th}$ percentile of $HD^{(i)}$. Algorithm \ref{algo_thrsld} describes the method of calculation of $C_{\alpha}$ through simulation.  Table \ref{tab_cutoff_circular} in the Appendix gives the values of $C_\alpha$ for 10\%, 5\%, and 1\% levels of significance for different values of $m$. 

\begin{algorithm}
    \caption{RCAG-Degree Distribution Test}
    \begin{algorithmic}[1]
        \State \textbf{Input:} A series of observations $\theta_1,\theta_2, \dots, \theta_{m}$.
        \If {$m=2n$}
        \State  Build an RCAG with $n$ vertices where $A_j = [\theta_{2j-1},\theta_{2j}]$, $1\leq j \leq n$.
	\State Compute $\hat{F}_n$, the empirical degree distribution for the RCAG obtained above. 
	\State Compute $HD = \frac{1}{\sqrt{2}} \sum_{i=1}^n \left( \sqrt{\hat{F}_n(\frac{i}{n}) -\hat{F}_n(\frac{i-1}{n})} - \sqrt{F^*(\frac{i}{n})-F^*(\frac{i-1}{n})}\right)^2$, where
    $F^*$ is the theoretical degree distribution for an RCAG with $n$ vertices.
        \State Reject null hypothesis of randomness if $HD>C_\alpha$, where $C_\alpha$ is computed using Algorithm \ref{algo_thrsld}.
        \ElsIf {$m = 2n+1$}
        \State Group observations into 2 groups $G_1=\{\theta_1, \ldots, \theta_{2n}\}$, $G_2= \{\theta_2, \ldots, \theta_{2n+1}\}$ of size $2n$ each.  
        \State Apply Steps 2-6 separately to groups $G_1$ and $G_2$ to obtain two test statistics $HD_1$ and $HD_2$.
        \State Reject null hypothesis of randomness if either of $HD_1$ or $HD_2$ is greater than $C_{\alpha/2}$.
        \EndIf
    \end{algorithmic}
    \label{algo_RCAG_DD}
\end{algorithm}

\begin{algorithm}
    \caption{Threshold Calculation}
    \begin{algorithmic}[1]
        \State Generate $k$ random series of length $2n$ from the Circular Uniform distribution (in this paper $k=1000$).
	\State Compute test statistic values $HD^{(1)}, HD^{(2)}, \dots, HD^{(k)}$ for each of these generated random series.
       \State Define threshold value $C_\alpha$ to be the $ 100(1-\alpha)^{th}$ percentile of $HD^{(i)}$'s for the level of significance $\alpha$.
    \end{algorithmic}
    \label{algo_thrsld}
\end{algorithm}

\section{Numerical Studies}
\label{exp}


In this section, we conduct experimental studies to examine the performance of our proposed RCAG-EP and RCAG-DD tests.
We use two time series models for circular data introduced in \cite{fisher_lee}. The first one is the \textit{linked autoregressive moving average} (LARMA) process, which is constructed as follows. Let $X_t$ be a stationary process on the real line and $\mu \in [0,2\pi)$ then the process $\Theta_t = g(X_t) + \mu \mod 2\pi$ is called the \textit{linked circular process} where $g : \mathbb{R} \rightarrow (-\pi, \pi)$ is an odd strictly monotonically increasing function known as link function. If $X_t$ is taken to be the ARMA(p,q) process, then $\Theta_t$ is said to follow the LARMA(p,q) process. The second one is the \textit{circular autoregressive} process of order $p$ (CAR(p)), which is a von Mises-based analogue of the AR(p) process for linear data. Here the conditional distribution of $\Theta_t$ is von Mises($\mu_t,\kappa$)  where $\mu_t = \mu + g\left( \alpha_1 g^{-1}(\theta_{t-1} - \mu \right) + \dots + \alpha_p g^{-1}(\theta_{t-p} -\mu))$ for given $\theta_{t-1}, \theta_{t-2},  \dots, \theta_{t-p}$ where $g$ is a link function. In this paper, we take $g(x) = 2\tan^{-1}x$.

We begin by providing examples that demonstrate the working of the RCAG-EP and RCAG-DD tests. First, for the RCAG-EP test, we simulate a random series from a circular uniform distribution. We apply the RCAG-EP test to the data and determine whether our test can detect the randomness in it, as outlined below

\begin{figure}[t]
    \centering
    \includegraphics[scale=0.6]{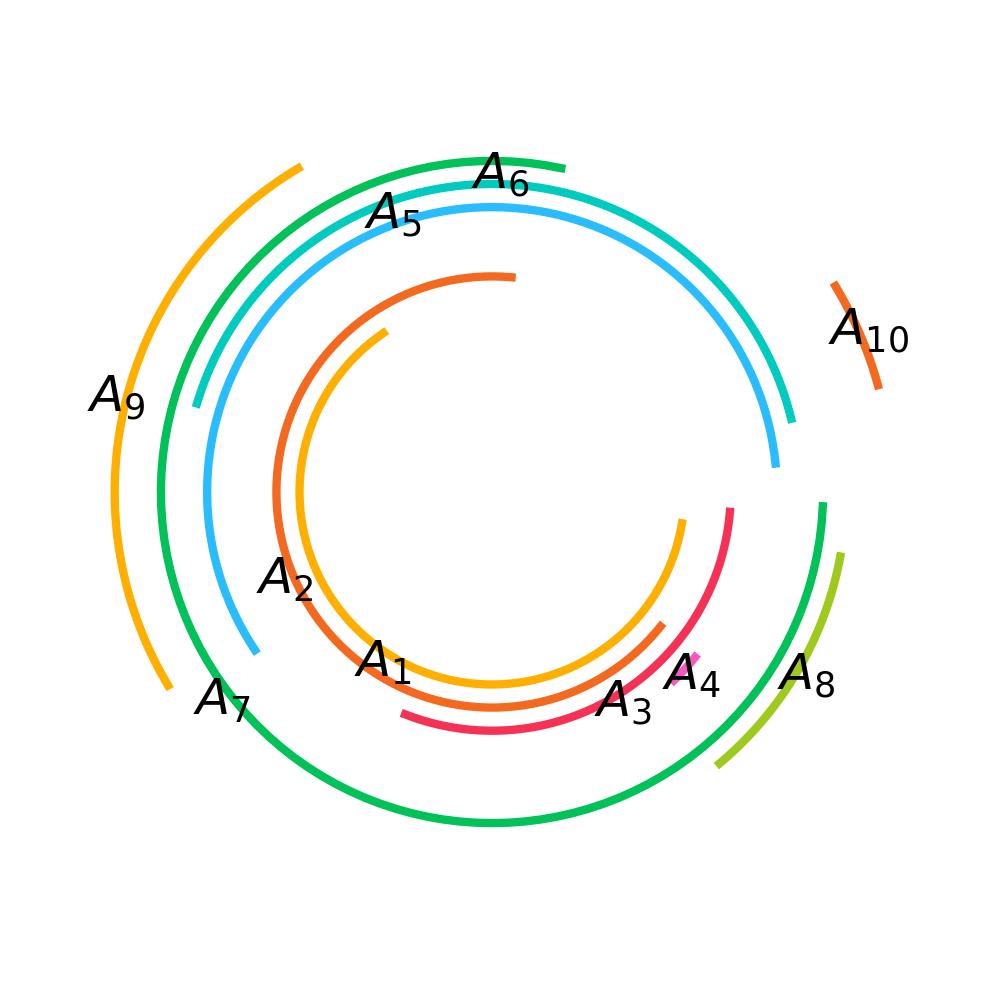}
    \caption{Arcs formed from a set of 20 random numbers as obtained in Example \ref{exm_RCAG_EP}}
    \label{fig_arcs}
\end{figure}

\begin{exmp}
\label{exm_RCAG_EP}
    Consider a set of 20 randomly generated observations (in radians) from the circular uniform distribution as
    $2.17, 6.12, 1.48, 5.61, 4.34, 6.20, 5.60, 5.48, 3.73, 0.10,$ $ 0.24, 2.85,$ $6.24, 1.36,$ $6.10, 5.41, 2.11, 3.68, 0.54, 0.27$.
    We apply the RCAG-EP test on it as given in Algorithm \ref{algo_RCAG_EP}. This gives us an RCAG with ten vertices corresponding to arcs $A_1 = [2.17, 6.12], A_2 = [ 1.48, 5.61], A_3 = [ 4.34, 6.20], A_4 = [ 5.60, 5.48],$ $ A_5 =  [ 3.73, 0.10], A_6 = [ 0.24, 2.85], A_7 = [ 6.24, 1.36], A_8 = [ 6.10, 5.41], A_9 = [ 2.11, 3.68], A_{10} = [ 0.54, 0.27]$ as shown in Figure \ref{fig_arcs}. Let us form five random pairs of arcs from these as $(A_8, A_7),(A_4, A_3), $ $(A_9, A_2), (A_1, A_{10}),  (A_5, A_6).$ Here, since the arcs in the pair $(A_1, A_{10})$ are non-intersecting, we get $\sum_{i=1}^{5}y_i = 1$. Now using an exact randomized test for $n=5$ as described in Section 1, the values of parameters obtained are $K_1 = \emptyset, K_2 =\{0,5\}$, and $\gamma = 0.368$ for $\alpha = 0.05$. Since $\sum_{i=1}^{5}y_i \notin K_1, K_2$,  we get, $\varphi(y_1,\dots,y_5) = 0$. Thus, we do not reject the null hypothesis of randomness at the 5\% level of significance. 
\end{exmp}

Next, to understand the working of the RCAG-DD test, consider the following example of 1000 observations. We consider the LAR(1) series to incorporate the dependence in the data.

\begin{exmp}
    Consider three different sets of 1000 observations generated (i) randomly from Circular Uniform distribution, (ii) randomly from von Mises(0,2) distributions, and (iii) from LAR(1) process with $\rho = 0.9$. We applied the RCAG-DD test as given in Algorithm \ref{algo_RCAG_DD}
    to all three sets of observations. Figure \ref{fig_dd_exmpl} 
    shows all degree distributions obtained. Here, when the observations were randomly generated from the Circular-Uniform and von Mises distributions, the empirical degree distribution looks similar to the theoretical degree distribution. However, when the observations are coming from a dependent series like LAR(1), the empirical degree distribution does not resemble the theoretical degree distribution. Specifically, the Hellinger distances between the theoretical and empirical pmfs for cases (i), (ii), and (iii) are 0.2082, 0.2354, and 2.1475, respectively. From Algorithm \ref{algo_thrsld}, the cutoff value for the 5\% level of significance is 0.7242. Hence, we do not reject the null hypothesis at the 5\% level of significance in cases (i) and (ii). Whereas we reject the null hypothesis at the 5\% level of significance when data is generated from the LAR(1) series.
\end{exmp}

To demonstrate the versatility of our proposed tests across different distributions and stochastic processes, we conduct the first simulation experiment as follows. We generate a series of 1000 observations from four different sources: (i) Circular Uniform distribution, (ii) von Mises distributions, (iii) Wrapped Cauchy distribution, and (iv) LAR(1) process. For each series, we construct an RCAG and assess randomness at the 5\% level of significance using both the RCAG-EP and RCAG-DD tests.
The results indicate that when observations are truly random, as in cases (i), (ii), and (iii), the tests do not reject the null hypothesis of randomness. However, for case (iv), where dependence is present, the tests successfully detect and reject the null hypothesis of randomness. This demonstrates the effectiveness of our approach in identifying dependence within data.

\begin{table}
    \caption{Power of RCAG-EP and RCAG-DD tests against LAR(1) process}
    \centering
    \hspace*{-1cm}
    \begin{tabular}{cccccccccccc}
        \hline
        & $\rho$ & 0.9 & 0.8 & 0.7 & 0.6 & 0.5 & 0.4 & 0.3 & 0.2 & 0.1 & 0 \\ 
        \hline
        \multirow{2}{5em}{$m=1000$} & RCAG-EP & 60.0 & 41.5 & 25.7 & 21.9 & 12.0 & 11.0 & 7.4 & 5.2 & 4.7 & 4.6 \\
        & RCAG-DD & 100.0 & 100.0 & 99.2 & 83.1 & 52.7 & 26.9 & 15.2 & 11.3 & 6.5 & 5.3\\ 
        \hline
        \multirow{2}{5em}{$m=2000$} & RCAG-EP & 87.5 & 71.6 & 52.8 & 37.9 & 23.0 & 16.9 & 9.7 & 7.6 & 5.4 & 5.2 \\
        & RCAG-DD &  100.0 & 100.0 & 100.0 & 95.1 & 67.5 & 31.9 & 16.1 & 8.6 & 5.4 & 5.1 \\ 
        \hline
        \multirow{2}{5em}{$m=8000$} & RCAG-EP & 100.0 & 99.9 & 98.1 & 88.8 & 68.6 & 42.6 & 23.3 & 10.6 & 6.4 & 6.3 \\
        & RCAG-DD & 100.0 & 100.0 & 100.0 & 100.0 & 99.0 & 72.3 & 33.7 & 14.4 & 8.7 & 5.7\\
        \hline
    \end{tabular}
    \hspace*{-1cm}
    \label{tab_LAR1}
\end{table}

To evaluate the power of the RCAG-EP and RCAG-DD tests, we conducted a series of simulation experiments. We begin by analyzing the LAR(1) process, generating a set of $m$ observations for varying values of $\rho$ with $m=1000,2000$ and $8000$. Both tests were applied at the 5\% level of significance, and their performance was assessed by repeating the procedure 1000 times and recording the frequency of rejection of the null hypothesis. The results are summarized in Table \ref{tab_LAR1}. Table \ref{tab_LAR1} demonstrates that the RCAG-DD test consistently outperforms the RCAG-EP test across all values of $m$ and $\rho$. The difference is significant, especially for smaller values of $m$ and $\rho$. As $\rho$ increases, indicating stronger dependence in the series, the power of both tests improves. At $\rho=0$, the rejection rates are close to the nominal level of significance, as expected. Additionally, the power of both tests increases with larger sample sizes.

\begin{figure}[t]
    \centering
    \includegraphics[width=0.98\linewidth]{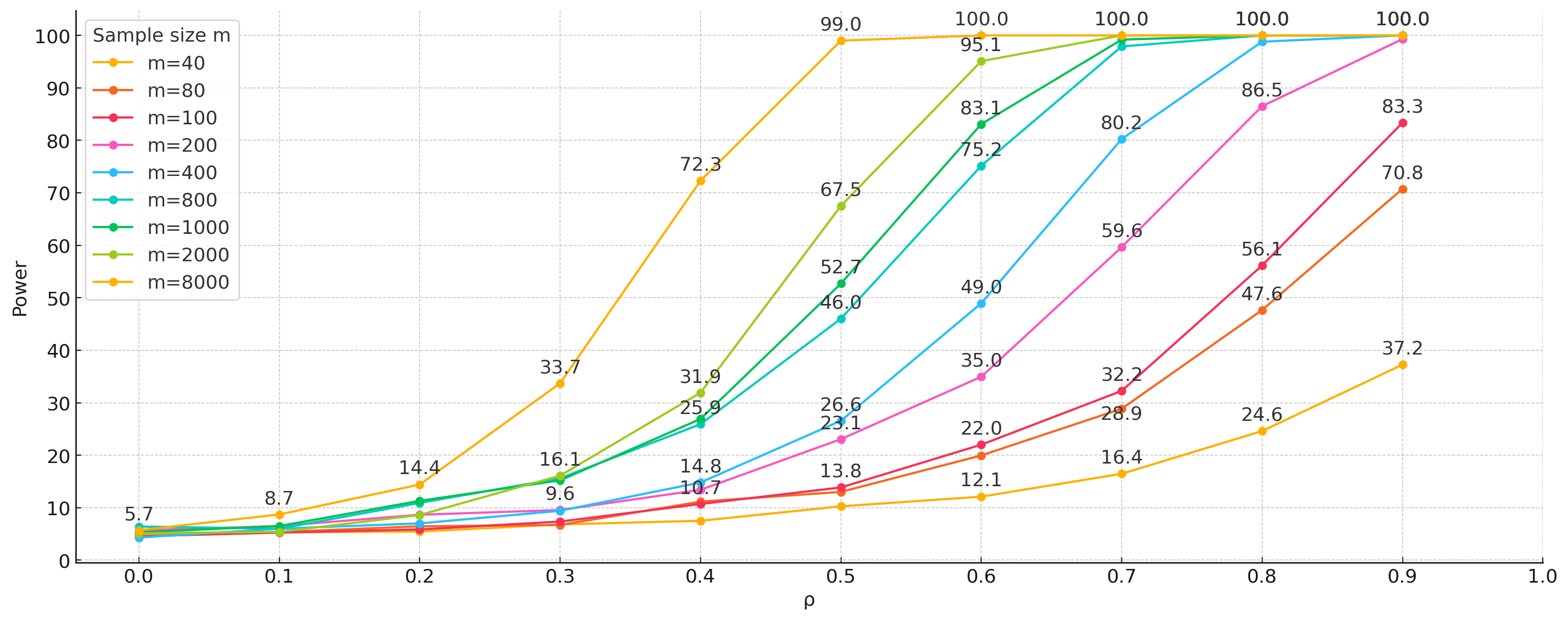}
    \caption{ Variation in power of RCAG-DD test against LAR(1) process for different values of $m$ and $\rho$.}
    \label{fig:LAR1}
\end{figure}

Since the RCAG-DD test exhibits superior performance, detecting dependence more than 99\% of the time for strong dependence ($\rho\geq 0.7)$ even with $m=1000$, we further examine its effectiveness for smaller values of $m$ to gain deeper insights into its performance. To do so, we conduct a similar experiment by generating 2000 series of observations from the LAR(1) process of size $m$ starting with $m=40$ and gradually increasing it to 800. Figure \ref{fig:LAR1} illustrates the variation in power for different values of $m$ and $\rho$. The results indicate that even with smaller sample sizes, the power of the RCAG-DD test improves as $\rho$ increases. Additionally, the test's power grows with larger sample sizes. For example, at $\rho=0.9$, the power rises from 37.2\% for $m=40$ to nearly 100\% for $m=200$, demonstrating its efficiency in identifying dependence even in moderately sized datasets. Moreover, the test maintains the desired empirical size across all values of $m$, with rejection rates at $\rho=0$ staying close to the 5\% level of significance.

To further evaluate the ability of the RCAG-EP and RCAG-DD tests to detect dependence beyond lag 1, we extend our analysis to the LAR(2) and LMA(1) processes. For each process, We generate 1000 series of length $m$, varying $\Vec{\rho}= (\rho_1,\rho_2)$ for LAR(2) with $m=200,2000,3000$ and $\rho$ for LMA(1), with $m=400,2000,3000$. The tests are then applied to these datasets. The results, summarized in Tables \ref{tab_LAR2} and \ref{tab_LMA1}, confirm that both RCAG-EP and RCAG-DD successfully detect such dependence in the data. Notably, the RCAG-DD test continues to outperform the RCAG-EP test. In the case of LMA(1) process, the power patterns resemble those observed for the LAR(1) process. 
For the LAR(2) process, as the value of $||\Vec{\rho}|| = \left(\rho_1^2 + \rho_2^2\right)^{1/2}$ increases, the power of both tests improves. Additionally, when $\rho_1$  is larger, the rejection rate is higher, suggesting that $\rho_1$  plays a more dominant role in driving dependence. These findings validate the reliability of RCAG-EP and RCAG-DD tests in detecting dependence across a range of LAR and LMA processes.

\begin{table}
    \caption{Power of RCAG-EP and RCAG-DD tests against LAR(2) process}
    \centering
    \begin{tabular}{cccccccc}
        \hline
        & $\Vec{\rho}$ & (0.5,0.4)  & (0.4,0.5) & (0.4,0.2) & (0.2,0.4)& (0.2,0.2) &  (0,0) \\ 
        \hline
         \multirow{2}{5em}{$m=200$} & RCAG-EP & 14.8 & 15.4 & 9.1 & 7.1 & 6.5 & 5.0\\
        & RCAG-DD & 86.3 & 75.5 & 23.5 & 12.9 & 11.6 & 5.1 \\ 
        \hline
        \multirow{2}{5em}{$m=2000$} & RCAG-EP & 75.6 & 65 & 25.1 & 12.4 & 9.0 & 5.7\\
        & RCAG-DD & 100.0 & 100.0 & 65.1 & 22.3 & 11.4 & 4.8 \\ 
        \hline
        \multirow{2}{5em}{$m=3000$} & RCAG-EP & 87.7& 81.4 & 29.8 & 15.2 & 9.1 & 4.2\\
        & RCAG-DD & 100.0 & 100.0 & 81.2 & 30.2 & 17.7 & 5.5\\ 
        \hline
    \end{tabular}
    \label{tab_LAR2}
\end{table}

\begin{table}
\caption{Power of RCAG-EP and RCAG-DD tests against LMA(1) process}
    \centering
    \hspace*{-0.9cm}
    \begin{tabular}{cccccccccccc}
        \hline
        & $\rho$ & 0.9 & 0.8 & 0.7 & 0.6 & 0.5 & 0.4 & 0.3 & 0.2 & 0.1 & 0 \\ 
        \hline
        \multirow{2}{5em}{$m=400$} & RCAG-EP & 10.1 & 11.3 & 10.2 & 9.5 & 8.3 & 7.4 & 6.1 & 7.4 & 6.9 & 5.9\\
        & RCAG-DD & 24.8 & 22.7 & 20.9 & 17.2 & 15.2 & 10.0 & 8.0 & 6.6 & 6.7 & 3.2\\ 
        \hline
        \multirow{2}{5em}{$m=2000$} & RCAG-EP & 22.6 & 22 & 20.4 & 18.2 & 17.4 & 13.1 & 10.9 & 7.8 & 5.7 & 5.1\\
        & RCAG-DD & 63.5 & 59.7 & 51.1 & 41.8 & 33.2 & 22.0 & 13.0 & 7.5 & 4.8 & 4.0 \\ 
        \hline
        \multirow{2}{5em}{$m=3000$} & RCAG-EP & 29.1 & 28.2 & 23.3 & 21.9 & 19.6 & 14.5 & 9.4 & 4.7 & 4.5 & 5.3\\
        & RCAG-DD &  80.1 & 74.6 & 67.3 & 57.3 & 43.9 & 32.1 & 19.1 & 10.1 & 6.5 & 6.1\\
        \hline
    \end{tabular}
    \hspace*{-0.9cm}
    \label{tab_LMA1}
\end{table}

Next, we evaluate the power of the tests against the CAR($p$) process. We take $p =2, \mu= 0 $ and $\alpha_1 = \alpha_2 = 0.5$ and vary the values of $\kappa$ and $m$. 
We generate 2000 series of $m$ observations from the CAR(2) series for different values of $\kappa$, starting with a small dataset of 40 observations and progressively increasing to larger datasets with up to 1500 observations. The simulation results, presented in Table \ref{tab_CAR2}, indicate that the power of both tests improves as the sample size increases. Additionally, we also find that power remains relatively stable across higher values of $\kappa$. Consistent with earlier findings, the RCAG-DD test outperforms the RCAG-EP test, demonstrating its superior ability to detect dependence, especially in smaller datasets. 

\begin{table}
\caption{Power of RCAG-EP and RCAG-DD tests against CAR($2$) process for $\mu= 0, \alpha_1 = \alpha_2 = 0.5$ for different values of $\kappa$ and $m$}
    \centering
    \begin{tabular}{ccccccccc}
        \hline
         & $m$ & 40 & 80 & 100 & 200 & 400 & 800 & 1500 \\ 
        \hline
        \multirow{4}{5em}{RCAG-EP}  & $\kappa=1$  &  9.9 & 9.1 & 10.5 & 9.5 & 16.9 & 18.5 & 29.9\\
        & $\kappa=3$  &  14.7 & 13.7 & 16.3 & 18.9 & 34.1 & 52.0 & 74.9\\
        & $\kappa=7$  & 14.6  & 14.1 & 18.5 & 23.8 & 43.2 & 65.2 & 88.1\\
        & $\kappa=13$ & 13.2 & 15.0 & 18.3 & 25.6 & 46.1 & 70.6 & 91.2\\ 
         \hline 
        \multirow{4}{5em}{RCAG-DD} & $\kappa=1$  & 12.1 &  18.1 & 26.9 & 35.4 & 53.5 & 79.4 & 95.1\\
        & $\kappa=3$  & 29.4 & 59.9 & 79.5 & 97.9 & 100.0 & 100.0 & 100.0\\
        & $\kappa=7$  &  30.8 & 69.3 & 86.3 & 99.5 & 100.0 & 100.0 & 100.0\\
        & $\kappa=13$ & 27.5 & 67.8 & 83.8 & 99.5 & 100.0 & 100.0 & 100.0\\
        \hline
    \end{tabular}
\label{tab_CAR2}
\end{table}

So far, our analysis has concentrated on datasets where the number of observations, $m$, is a multiple of four for the RCAG-EP test and a multiple of two for the RCAG-DD test. To further assess the applicability of our approach, we expanded our analysis to include cases where $m$ does not adhere to these constraints. This extension allows us to gain deeper insights into the performance and reliability of our tests under more varied and practical data conditions.

First, for the RCAG-EP test, we generate 1000 series of $m$ observations from a LAR(1) process with varying values of $\rho$. We consider four values of $m$: (i) $m = 2000$, (ii) $m = 2001$, (iii) $m = 2002$, and (iv) $m = 2003$. We apply the RCAG-EP test to each of these datasets, assessing randomness at the 5\% level of significance. For cases where $m$ is not a multiple of four, we implement Algorithm \ref{algo_RCAG_EP} using the Benjamini-Hochberg correction. The results of this experiment are presented in Table \ref{tab:not_multiple_of_4}. The simulation results show that the outcomes for cases (ii), (iii), and (iv) are in close agreement with those of case (i), confirming that the method continues to perform well even when $m$ is not a multiple of four. 

\begin{table}[t]
\caption{Comparison of performance of RCAG-EP Test against LAR(1) process when $m$ is and is not a multiple of four.}
    \centering
    \begin{tabular}{ccccccccccc}
        \hline
        $\rho$ & 0.9 & 0.8 & 0.7 & 0.6 & 0.5 & 0.4 & 0.3 & 0.2 & 0.1 & 0 \\
        \hline
        $m=2000$ &  87.5 & 71.6 & 52.8 & 37.9 & 23 & 16.9 & 9.7 & 7.6 & 5.4 & 5.2 \\
        \hline
        $m=2001$
         & 97.9 & 86.4 &65.9 & 45.9 & 27.8 & 19.8 & 12.7 & 8.8 & 5.3 & 4.2\\
        \hline
        $m=2002$
        &  98.5 & 91.2 & 71.3 & 48.7 & 32.3 & 19.7 & 12 & 9.3 & 5.9 & 4.7\\
        \hline
        $m=2003$
        & 99.1 & 90 & 72.4 & 54.4 & 32.8 & 19 & 11.1 & 8.7 & 5.4 & 5.6 \\
        \hline
    \end{tabular}
     \label{tab:not_multiple_of_4}
\end{table}

For the RCAG-DD test, we generate 1000 series of $m$ observations from the LAR(1) process with different values of $\rho$ when $m = 1000$ and $m = 1001$. For each dataset, we apply the RCAG-DD test as outlined in Algorithm \ref{algo_RCAG_DD}, using the 5\% level of significance to assess randomness. As shown in Table \ref{tab:not_multiple_of_2}, the test exhibits similar power for both values of $m$. These results confirm that the method remains effective even when the number of observations is not a multiple of two.

\begin{table}
 \caption{Comparison of performance of RCAG-DD Test against LAR(1) process when $m$ is and is not a multiple of two.}
    \centering
    \begin{tabular}{ccccccccccc}
        \hline
        $\rho$ & 0.9 & 0.8 & 0.7 & 0.6 & 0.5 & 0.4 & 0.3 & 0.2 & 0.1 & 0 \\
        \hline
        $m=1000$ & 100 & 100 & 99.2 & 83.1 & 52.7 & 26.9 & 15.2 & 11.3 & 6.5 & 5.3\\ 
        \hline
        $m=1001$  & 100 & 100 & 100 & 91.7 & 61.7 & 29.9 & 16.5 & 10.2 & 7.8 & 4.3 \\
        \hline
    \end{tabular}
    \label{tab:not_multiple_of_2}
\end{table}

Finally, we assessed the performance of our tests against permuted sequences of random series. For this, we generated a set of 1000 random observations from (i) the circular uniform distribution and (ii) the von Mises (0,2) distribution and created 1000 permutations for each. We check the randomness for each of these permuted series at the 5\% level of significance. The RCAG-EP test rejected the null hypothesis for 5.2\% permutations from (i) and 5.9\% permutations from (ii). The RCAG-DD test showed rejection rates of 4.7\% and 5\% permutations from (i) and (ii), respectively. These results confirm the robustness of our tests under permutation scenarios.

\section{Real World Applications} 
\label{real_data}
\subsection{Wind direction data}
We analyze hourly wind direction data from Basel, Switzerland, collected between 01 January 2022 and 24 January 2023. The dataset consists of 9336 angular observations and is publicly available on Meteoblue, a meteorological platform developed by the University of Basel. Applying the RCAG-EP test as per Algorithm \ref{algo_RCAG_EP} on the data, we obtain $|\hat{p} - 1/6| = 0.1632$. Since this is greater than  $c = 0.0151$, we reject the null hypothesis of randomness at the 5\% level of significance.  Now, applying the RCAG-DD test as per Algorithm \ref{algo_RCAG_DD}, the Hellinger distance obtained between the empirical and theoretical pmfs is 0.7998, which is greater than the 5\% level of significance cutoff of 0.3520 obtained using Algorithm \ref{algo_thrsld}. Consequently, we again reject the null hypothesis of randomness.
Thus, both the RCAG-EP and RCAG-DD tests indicate the presence of dependence in the wind direction data. 

\subsection{Location data of fireball showers, meteor falls, and craters on Earth’s surface}
We analyze three datasets of extraterrestrial objects impacting the Earth’s surface. These datasets include the locations (latitude and longitude) of: (i) craters formed by meteor impacts, (ii) reported fireball shower events, and (iii) meteorite falls \cite{fireball_data}.
Since latitude and longitude are defined within $[-\pi/2, \pi/2]$ and $[-\pi,\pi]$, respectively, we scale the data to $[0,2\pi]$ to standardize the range. We then apply the RCAG-EP and RCAG-DD tests separately to the latitude and longitude datasets, testing each for randomness independently, as their distributions are fitted separately in \cite{fireball}.

The first dataset of the location of craters contains $m=70$ angular data points. By applying the RCAG-EP test on the data as given in Algorithm \ref{algo_RCAG_EP}, we get the adjusted p-values for the latitude dataset as $0.1239, 0.0511 $ and $0.6644$, and the longitude dataset as $0.2781,0.0511,$ and $0.1240$. 
Since all exceed the 5\% level of significance, the null hypothesis of randomness is not rejected for either dataset.
For the RCAG-DD test, the Hellinger distances between the theoretical and empirical degree distributions are 0.4918 for the latitude dataset and 0.5052 for the longitude dataset, both below the threshold of 0.5684 at the 5\% level of significance, again indicating no evidence against randomness.

The second dataset contains $m=769$ angular observations of the location of the fireball showers. The RCAG-EP yields adjusted p-values below $0.0001$ for both latitude and longitude, indicating significant deviation from randomness. Applying the RCAG-DD test, the test statistics obtained are 0.9026 and 0.9026 for latitude data and 0.3645 and 0.3949 for longitude data. Since at least one of these exceeds the threshold of 0.3940 for both datasets, we reject the null hypothesis at the 5\% level of significance. 

The final dataset includes $m=17069$ angular observations of meteorite fall locations. The RCAG-EP test returns adjusted p-values below $0.0001$ for both latitude and longitude, indicating the presence of non-randomness in the data.  For the RCAG-DD test, the test statistics obtained are 0.7566 and 0.7333 for latitude data and 0.7308 and 0.7178 for longitude data. As test statistics exceed the threshold of 0.3947 for both datasets, we reject the null hypothesis at the 5\% level of significance. 

Our findings indicate that crater locations exhibit randomness, while fireball showers and meteor falls do not. Since Watson's and Rao's spacing tests rely on the assumption of randomness, applying these tests is valid for the crater data but not for fireball and meteor data. Thus, the conclusions made in \cite{fireball} about the fireball shower and meteorite fall data following a non-uniform, yet not von Mises distribution, are possibly incorrect, as the observed deviations from a von Mises distribution are due to the non-random nature of the data.
Instead of fitting von Mises or similar distributions, future analyses should consider models that account for underlying dependencies, such as spatiotemporal models. Such approaches would better capture the inherent structure and non-random patterns in the data, leading to more accurate interpretations of extraterrestrial object impacts on Earth.

\section{Concluding Remarks}
\label{conclusions}

In this paper, we introduced novel tests for assessing randomness in circular data through the concept of RCAGs. Specifically, we developed the RCAG-EP test and the RCAG-DD test, leveraging two key properties of RCAGs, namely, edge probability and vertex degree distribution. These tests are constructed based on the principle that these graph-based properties are independent of the underlying distribution of the data, depending solely on the assumption of independence. This distribution-free characteristic makes the tests broadly applicable across diverse data types and domains.

Through extensive simulation studies, we demonstrated the robustness and effectiveness of our methods under various settings. Both tests effectively detected dependencies in both LARMA and CAR processes. We further validated the practical relevance of our approach through real-world applications, including datasets on wind direction and extraterrestrial object locations. The results highlighted the test's ability to uncover non-randomness in diverse and complex data. 

Future research can build upon this work by exploring additional structural properties of RCAGs, such as the chromatic number, independence number, and maximum clique size. Investigating these aspects may not only deepen the theoretical understanding of RCAGs but also uncover new use cases in graph-based statistical analysis. Another promising direction is extending this methodology to higher-dimensional spaces, such as spheres, torus, and hyperspheres. Developing randomness tests in such settings could enable the study of more intricate datasets encountered in fields like biology, astrophysics, geospatial analysis, and machine learning.

\section{Data Availability Statement}

The hourly wind direction data of Basel, Switzerland, is openly available on Meteoblue at 
https://www.meteoblue.com/en/weather/archive/export. \\
The location data for extraterrestrial objects can be accessed on Harvard Dataverse at https://doi.org/10.7910/DVN/FLNQM5 \citep{fireball_data}.

\appendix

\section{Threshold Values for the RCAG-DD Test}

Table \ref{tab_cutoff_circular} presents the threshold values ($C_\alpha$) for the RCAG-DD test applied on $m$ observations at the level of significance $\alpha$, as computed using Algorithm \ref{algo_thrsld}.

\begin{table}[t]
\caption{Threshold values for RCAG-DD test at different levels of significance with $m$ observations.}
    \centering
    \begin{tabular}{cccc||cccc}
        \hline
        \multirow{2}{*}{$m$} & \multicolumn{3}{c||}{Level of Significance} & \multirow{2}{*}{$m$} & \multicolumn{3}{c}{Level of Significance} \\
        & 10\%  & 5\%  & 1\% &  & 10\%  & 5\%  & 1\% \\
        \hline
        40   & 0.61479 & 0.63671 & 0.71622 & 1200 & 0.37465 & 0.37881 & 0.38782 \\
        60   & 0.55796 & 0.57901 & 0.61061 & 1500 & 0.36822 & 0.37237 & 0.37968 \\
        80   & 0.52982 & 0.54835 & 0.58169 & 2000 & 0.36317 & 0.36693 & 0.37196 \\
        100  & 0.50929 & 0.52625 & 0.56101 & 2300 & 0.35944 & 0.36258 & 0.36697 \\
        200  & 0.45365 & 0.46431 & 0.48605 & 2700 & 0.35640 & 0.35988 & 0.36477 \\
        300  & 0.43288 & 0.44144 & 0.46347 & 3000 & 0.35453 & 0.35735 & 0.36214 \\
        400  & 0.41672 & 0.42533 & 0.44297 & 3300 & 0.35353 & 0.35639 & 0.36121 \\
        500  & 0.40645 & 0.41239 & 0.42510 & 3700 & 0.35162 & 0.35443 & 0.36130 \\
        600  & 0.39746 & 0.40262 & 0.41657 & 4000 & 0.35037 & 0.35249 & 0.35696 \\
        700  & 0.39256 & 0.39836 & 0.41047 & 5000 & 0.34854 & 0.35102 & 0.35613 \\
        800  & 0.38662 & 0.39131 & 0.40196 & 6000 & 0.34642 & 0.34827 & 0.35230 \\
        900  & 0.38347 & 0.38864 & 0.39882 & 7000 & 0.34457 & 0.34639 & 0.34955 \\
        1000 & 0.37982 & 0.38380 & 0.39133 & 8000 & 0.34365 & 0.34519 & 0.34817 \\
        \hline
    \end{tabular}
    \label{tab_cutoff_circular}
\end{table}

\bibliography{reference}

\begin{thebibliography}{}

\bibitem[Ajne, 1968]{Ajne}
Ajne, B. (1968).
\newblock A simple test for uniformity of a circular distribution.
\newblock {\em Biometrika}, 55(2):343--354.

\bibitem[Batschelet, 1981]{Batschelet}
Batschelet, E. (1981).
\newblock {\em Circular statistics in biology}.
\newblock Academic Press.

\bibitem[Benjamini and Yekutieli, 2001]{BY_correction}
Benjamini, Y. and Yekutieli, D. (2001).
\newblock The control of the false discovery rate in multiple testing under dependency.
\newblock {\em Annals of statistics}, pages 1165--1188.

\bibitem[Blimpo, 2014]{edu_Africa}
Blimpo, M.~P. (2014).
\newblock Team incentives for education in developing countries: A randomized field experiment in benin.
\newblock {\em American Economic Journal: Applied Economics}, 6(4):90--109.

\bibitem[Brunsdon and Corcoran, 2006]{so_sc1}
Brunsdon, C. and Corcoran, J. (2006).
\newblock Using circular statistics to analyse time patterns in crime incidence.
\newblock {\em Computers, Environment and Urban Systems}, 30(3):300--319.

\bibitem[Caires and Wyatt, 2003]{ocean}
Caires, S. and Wyatt, L. (2003).
\newblock A linear functional relationship model for circular data with an application to the assessment of ocean wave measurements.
\newblock {\em Journal of agricultural, biological, and environmental statistics}, 8:153--169.

\bibitem[Cuckler and Kahn, 2009]{ham_dirac}
Cuckler, B. and Kahn, J. (2009).
\newblock Hamiltonian cycles in dirac graphs.
\newblock {\em Combinatorica}, 29:299--326.

\bibitem[Damaschke, 1993]{2}
Damaschke, P. (1993).
\newblock Paths in interval graphs and circular arc graphs.
\newblock {\em Discrete Mathematics}, 112(1-3):49--64.

\bibitem[De~la Fuente~Marcos and De~la Fuente~Marcos, 2015]{laFuenteMarcos}
De~la Fuente~Marcos, C. and De~la Fuente~Marcos, R. (2015).
\newblock Recent multi-kiloton impact events: are they truly random?
\newblock {\em Monthly Notices of the Royal Astronomical Society: Letters}, 446(1):L31--L35.

\bibitem[Diethe et~al., 2015]{ml1}
Diethe, T., Twomey, N., and Flach, P. (2015).
\newblock Bayesian modelling of the temporal aspects of smart home activity with circular statistics.
\newblock In {\em Joint European Conference on Machine Learning and Knowledge Discovery in Databases}, pages 279--294. Springer.

\bibitem[Dirac, 1952]{dirac}
Dirac, G.~A. (1952).
\newblock Some theorems on abstract graphs.
\newblock {\em Proceedings of the London Mathematical Society}, 3(1):69--81.

\bibitem[Dur{\'a}n et~al., 2014]{CAG_S2}
Dur{\'a}n, G., Grippo, L.~N., and Safe, M.~D. (2014).
\newblock Structural results on circular-arc graphs and circle graphs: a survey and the main open problems.
\newblock {\em Discrete Applied Mathematics}, 164:427--443.

\bibitem[Fern{\'a}ndez-Dur{\'a}n, 2007]{bio1}
Fern{\'a}ndez-Dur{\'a}n, J. (2007).
\newblock Models for circular--linear and circular--circular data constructed from circular distributions based on nonnegative trigonometric sums.
\newblock {\em Biometrics}, 63(2):579--585.

\bibitem[Fisher, 1995]{Fisher}
Fisher, N.~I. (1995).
\newblock {\em Statistical analysis of circular data}.
\newblock cambridge university press.

\bibitem[Fisher and Lee, 1992]{ecology1}
Fisher, N.~I. and Lee, A.~J. (1992).
\newblock Regression models for an angular response.
\newblock {\em Biometrics}, pages 665--677.

\bibitem[Fisher and Lee, 1994]{fisher_lee}
Fisher, N.~I. and Lee, A.~J. (1994).
\newblock Time series analysis of circular data.
\newblock {\em Journal of the Royal Statistical Society: Series B (Methodological)}, 56(2):327--339.

\bibitem[Garc{\'\i}a-Portugu{\'e}s et~al., 2015]{bivar5}
Garc{\'\i}a-Portugu{\'e}s, E., Crujeiras, R.~M., and Gonz{\'a}lez-Manteiga, W. (2015).
\newblock Central limit theorems for directional and linear random variables with applications.
\newblock {\em Statistica Sinica}, pages 1207--1229.

\bibitem[Garc{\'\i}a-Portugu{\'e}s and Verdebout, 2018]{GPG}
Garc{\'\i}a-Portugu{\'e}s, E. and Verdebout, T. (2018).
\newblock An overview of uniformity tests on the hypersphere.
\newblock {\em arXiv preprint arXiv:1804.00286}.

\bibitem[Garey and Johnson, 2002]{np_hard}
Garey, M.~R. and Johnson, D.~S. (2002).
\newblock {\em Computers and intractability}, volume~29.
\newblock wh freeman New York.

\bibitem[Ghosh et~al., 2023]{fireball_data}
Ghosh, P., Chatterjee, D., and Banerjee, A. (2023).
\newblock {A Short Note on Directional Statistical Distribution of Fireball Meteor Shower and Meteor Crater on the Earth's Surface}.

\bibitem[Ghosh et~al., 2024]{fireball}
Ghosh, P., Chatterjee, D., and Banerjee, A. (2024).
\newblock On the directional nature of celestial object’s fall on the earth (part 1: distribution of fireball shower, meteor fall, and crater on earth’s surface).
\newblock {\em Monthly Notices of the Royal Astronomical Society}, 531(1):1294--1307.

\bibitem[Goette et~al., 2012]{group_ties}
Goette, L., Huffman, D., and Meier, S. (2012).
\newblock The impact of social ties on group interactions: Evidence from minimal groups and randomly assigned real groups.
\newblock {\em American Economic Journal: Microeconomics}, 4(1):101--115.

\bibitem[Hanbury, 2003]{comp_vision1}
Hanbury, A. (2003).
\newblock Circular statistics applied to colour images.
\newblock In {\em 8th Computer Vision Winter Workshop}, volume 91, No. 1-2, pages 53--71. Citeseer.

\bibitem[Heams, 2014]{bio_random}
Heams, T. (2014).
\newblock Randomness in biology.
\newblock {\em Mathematical Structures in Computer Science}, 24(3):e240308.

\bibitem[Jammalamadaka and Sengupta, 2001]{SenGupta}
Jammalamadaka, S.~R. and Sengupta, A. (2001).
\newblock {\em Topics in circular statistics}, volume~5.
\newblock world scientific.

\bibitem[Jupp and Mardia, 1980]{cor2}
Jupp, P.~E. and Mardia, K.~V. (1980).
\newblock A general correlation coefficient for directional data and related regression problems.
\newblock {\em Biometrika}, 67(1):163--173.

\bibitem[Kubiak and Jonas, 2007]{psyc1}
Kubiak, T. and Jonas, C. (2007).
\newblock Applying circular statistics to the analysis of monitoring data.
\newblock {\em European Journal of Psychological Assessment}, 23(4):227--237.

\bibitem[Kuiper, 1960]{Kuiper}
Kuiper, N.~H. (1960).
\newblock Tests concerning random points on a circle.
\newblock In {\em Nederl. Akad. Wetensch. Proc. Ser. A}, volume 63, No. 1, pages 38--47.

\bibitem[Landler et~al., 2021]{bio3}
Landler, L., Ruxton, G.~D., and Malkemper, E.~P. (2021).
\newblock Advice on comparing two independent samples of circular data in biology.
\newblock {\em Scientific reports}, 11(1):20337.

\bibitem[Lang et~al., 2020]{wind_dir}
Lang, M.~N., Schlosser, L., Hothorn, T., Mayr, G.~J., Stauffer, R., and Zeileis, A. (2020).
\newblock Circular regression trees and forests with an application to probabilistic wind direction forecasting.
\newblock {\em Journal of the Royal Statistical Society Series C: Applied Statistics}, 69(5):1357--1374.

\bibitem[Lin and Szwarcfiter, 2009]{CAG_S1}
Lin, M.~C. and Szwarcfiter, J.~L. (2009).
\newblock Characterizations and recognition of circular-arc graphs and subclasses: A survey.
\newblock {\em Discrete Mathematics}, 309(18):5618--5635.

\bibitem[Mardia and Puri, 1978]{bivar2}
Mardia, K.~V. and Puri, M.~L. (1978).
\newblock A spherical correlation coefficient robust against scale.
\newblock {\em Biometrika}, 65(2):391--395.

\bibitem[Morellato et~al., 2010]{bio2}
Morellato, L. P.~C., Alberti, L., and Hudson, I.~L. (2010).
\newblock {\em Applications of circular statistics in plant phenology: a case studies approach}.
\newblock Springer.

\bibitem[Prati et~al., 2008]{comp_vision2}
Prati, A., Calderara, S., and Cucchiara, R. (2008).
\newblock Using circular statistics for trajectory shape analysis.
\newblock In {\em 2008 IEEE Conference on Computer Vision and Pattern Recognition}, pages 1--8. IEEE.

\bibitem[Puri and Rao, 1977]{bivar3}
Puri, M.~L. and Rao, J. (1977).
\newblock Problems of association for bivariate circular data and a new test of independence.
\newblock {\em Multivariate Analysis IV, Ed. PR Krishnaiah}, pages 513--22.

\bibitem[Rao, 1976]{rao's_spacing}
Rao, J. (1976).
\newblock Some tests based on arc-lengths for the circle.
\newblock {\em Sankhy{\=a}: The Indian Journal of Statistics, Series B}, pages 329--338.

\bibitem[Rayleigh, 1919]{Rayleigh}
Rayleigh, L. (1919).
\newblock {XXXI. O}n the problem of random vibrations, and of random flights in one, two, or three dimensions.
\newblock {\em The London, Edinburgh, and Dublin Philosophical Magazine and Journal of Science}, 37(220):321--347.

\bibitem[Rivest et~al., 2016]{ecology2}
Rivest, L.-P., Duchesne, T., Nicosia, A., and Fortin, D. (2016).
\newblock A general angular regression model for the analysis of data on animal movement in ecology.
\newblock {\em Journal of the Royal Statistical Society Series C: Applied Statistics}, 65(3):445--463.

\bibitem[Rohatgi and Saleh, 2015]{Book_Rohatgi}
Rohatgi, V.~K. and Saleh, A. M.~E. (2015).
\newblock {\em An introduction to probability and statistics}.
\newblock John Wiley \& Sons.

\bibitem[Rothman, 1971]{bivar1}
Rothman, E.~D. (1971).
\newblock Tests of coordinate independence for a bivariate sample on a torus.
\newblock {\em The Annals of Mathematical Statistics}, pages 1962--1969.

\bibitem[S{\'a}rk{\"o}zy et~al., 2003]{Hamiltonian_number}
S{\'a}rk{\"o}zy, G.~N., Selkow, S.~M., and Szemer{\'e}di, E. (2003).
\newblock On the number of hamiltonian cycles in dirac graphs.
\newblock {\em Discrete Mathematics}, 265(1-3):237--250.

\bibitem[Scheinerman, 1988]{3}
Scheinerman, E.~R. (1988).
\newblock Random interval graphs.
\newblock {\em Combinatorica}, 8:357--371.

\bibitem[Shieh and Johnson, 2005]{bivar4}
Shieh, G.~S. and Johnson, R.~A. (2005).
\newblock Inferences based on a bivariate distribution with von mises marginals.
\newblock {\em Annals of the Institute of Statistical Mathematics}, 57:789--802.

\bibitem[Shieh et~al., 1994]{cor3}
Shieh, G.~S., Johnson, R.~A., and Frees, E.~W. (1994).
\newblock Testing independence of bivariate circular data and weighted degenerate u-statistics.
\newblock {\em Statistica Sinica}, pages 729--747.

\bibitem[Sides et~al., 2023]{ml2}
Sides, K., Kilungeja, G., Tapia, M., Kreidl, P., Brinkmann, B.~H., and Nasseri, M. (2023).
\newblock Analyzing physiological signals recorded with a wearable sensor across the menstrual cycle using circular statistics.
\newblock {\em Frontiers in Network Physiology}, 3:1227228.

\bibitem[Tsay, 2005]{time_series}
Tsay, R.~S. (2005).
\newblock {\em Analysis of financial time series}.
\newblock John wiley \& sons.

\bibitem[Vasileios, 2005]{4}
Vasileios, I. (2005).
\newblock {\em Random interval graphs}.
\newblock PhD thesis, Master’s thesis, University of Essex.

\bibitem[Watson and Beran, 1967]{cor1}
Watson, G. and Beran, R. (1967).
\newblock Testing a sequence of unit vectors for serial correlation.
\newblock {\em Journal of Geophysical Research}, 72(22):5655--5659.

\bibitem[Watson, 1961]{Watson}
Watson, G.~S. (1961).
\newblock Goodness-of-fit tests on a circle.
\newblock {\em Biometrika}, 48(1/2):109--114.

\bibitem[Zhan et~al., 2019]{cor4}
Zhan, X., Ma, T., Liu, S., and Shimizu, K. (2019).
\newblock On circular correlation for data on the torus.
\newblock {\em Statistical papers}, 60:1827--1847.

\end{thebibliography}

\end{document}